\documentclass[reprint,
 amsmath,amssymb,
 aps, physrev,
 superscriptaddress, 
]{revtex4-2}

\usepackage{graphicx}
\usepackage{dcolumn}
\usepackage{bm}
\usepackage{amssymb}
\usepackage{multirow}
\usepackage{gensymb}
\usepackage{xcolor}
\usepackage[colorlinks=true,linkcolor=blue,citecolor=blue,urlcolor=blue]{hyperref}

\begin{document}

\preprint{APS/123-QED}

\title{\textbf{Disentangling magnetic and optical contributions in ultrafast dynamics of antiperovskite non-collinear antiferromagnets} 
}%

\author{Jozef Kimák}
\author{Tomáš Ostatnický}%
\author{Markéta Nerodilová}%
\affiliation{%
Faculty of Mathematics and Physics, Charles University, Ke Karlovu 3, 121 16,  Prague 2, Czech Republic}%

\author{Freya Johnson}
\affiliation{
Cavendish Laboratory, University of Cambridge, JJ Thomson Ave, CB3 0HE, Cambridge, United Kingdom
}%

\author{Ondřej Faiman}%
\author{Tomáš Trejtnar}%
\affiliation{%
Faculty of Mathematics and Physics, Charles University, Ke Karlovu 3, 121 16,  Prague 2, Czech Republic}%

\author{David Boldrin}
\author{Frederic Rendell-Bhatti}
\affiliation{
SUPA, School of Physics and Astronomy, University of Glasgow, Glasgow, G12 8QQ, 
United Kingdom
}%

\author{Jan Zemen}
\affiliation{
Faculty of Electrical Engineering, Czech Technical University, Technická 2, Prague 6, 166 00, 
Czech Republic
}%

\author{Bin Zou}%
\author{Andrei P. Mihai}%
\affiliation{%
LoMaRe Technologies Ltd., 6 London Street, EC3R 7LP London, United Kingdom}%

\author{Xudong Sun}%
\author{Fengzhi Yu}%
\affiliation{%
Jiangsu LoMaRe Chip Technology Co., Ltd., 400 Hanjiang Road, Changzhou, China}%

\author{Eva Schmoranzerová}%
\author{Lukáš Nádvorník}%
\affiliation{%
Faculty of Mathematics and Physics, Charles University, Ke Karlovu 3, 121 16,  Prague 2, Czech Republic}%

\author{Lesley F. Cohen}
\affiliation{
Department of Physics, Blackett Laboratory, Imperial College London, London SW7 2AZ, United Kingdom
}%

\author{Petr Němec}%
 \email{Contact author: petr.nemec@matfyz.cuni.cz}
\affiliation{%
Faculty of Mathematics and Physics, Charles University, Ke Karlovu 3, 121 16,  Prague 2, Czech Republic}%

\date{\today}

\begin{abstract} Non-collinear antiferromagnets are a class of spin-polarized antiferromagnets in which chiral spin textures give rise to Berry-curvature-driven phenomena, such as the anomalous Hall effect (AHE), without net magnetization. We investigate the properties of thin films of antiperovskite non-collinear antiferromagnetic metals Mn$_3$NiN and Mn$_3$GaN using pump-probe experiments. In both materials, we observe a strong dependence of pump-polarization-independent dynamics, induced by femtosecond laser pulses, on the angle between the sample normal and the direction of probe propagation. In Mn$_3$NiN, where the presence of a sizable AHE indicates the $\Gamma^{4g}$ phase, the measured magneto-optical (MO) signals acquire an additional, strong dependence on the external magnetic field when the probe pulses are incident at nonzero angles. In contrast, in Mn$_3$GaN, where the absence of AHE indicates the $\Gamma^{5g}$ phase, the measured signals do not depend on the magnetic field. Using probe-polarization-resolved measurements combined with full optical modeling based on Yeh’s formalism, we quantitatively separate magnetic and non-magnetic contributions to the measured signals. We show that in Mn$_3$NiN, the observed magnetic field dependence results from field-controlled redistribution of magnetic domain populations, enabled by their piezomagnetic moments and detected by a Kerr-like MO effect, while this effect is absent in Mn$_3$GaN. Temperature-dependent measurements reveal a change from single-step to two-step quenching dynamics with increasing temperature in Mn$_3$NiN. This behavior contrasts with the nearly temperature-independent quenching dynamics reported for the non-collinear antiferromagnetic Heusler compound Mn$_3$Sn, but resembles the crossover from type-I to type-II demagnetization dynamics in metallic ferromagnets.
\end{abstract}

\maketitle


\section{\label{sec:level1} Introduction}

Antiferromagnets (AFs) have recently emerged as a highly promising class of materials for next-generation spintronic devices, with the potential to replace ferromagnets (FMs) as active elements \cite{Jungwirth2016AntiferromagneticSpintronics, Baltz2018AntiferromagneticSpintronics}. This interest is primarily motivated by their compensated magnetic structure, which suppresses stray magnetic fields and thereby enables higher device integration densities and smaller magnetic bits. In addition, characteristic frequencies of uniform spin precession in AFs can reach terahertz region \cite{Nemec2018AntiferromagneticOpto-spintronics}, offering the prospect of significantly faster information processing. However, in conventional collinear AFs, the absence of spin splitting in the electronic band structure prevents the generation of non-relativistic macroscopic spin currents \cite{Baltz2018AntiferromagneticSpintronics}, which are essential for many spintronic devices, including magnetic tunnel junctions (MTJs). Furthermore, this lack of spin splitting makes their experimental characterization challenging, since effects that are linear in magnetization \textbf{M}, such as the anomalous Hall effect (AHE) and the magneto-optical Kerr effect (MOKE), vanish in these systems \cite{Nemec2018AntiferromagneticOpto-spintronics, Jungwirth2016AntiferromagneticSpintronics}.

To overcome these limitations, increasing attention has been directed toward non-collinear AFs (nc-AFs), which combine key advantages of both FMs and collinear AFs \cite{Rimmler2024Non-collinearSpintronics, Guo2025SpinPolarizedSpintronics}. Owing to their non-collinear spin structures, these materials exhibit spin-split electronic bands while retaining a compensated magnetic structure. Mn-based coplanar three-sublattice nc-AFs can be divided into two main families according to their crystal and spin structures \cite{Rimmler2024Non-collinearSpintronics}. The first family comprises Heusler compounds of the form Mn$_3$X (X = Sn, Ge, Ga), which crystallize in a hexagonal lattice and exhibit a kagome-like spin arrangement within the (0001) plane. These systems have already demonstrated significant potential for spintronic applications, as evidenced by the experimental observation of AHE \cite{Nakatsuji2015LargeTemperature, Nayak2016LargeMn3Ge}, anomalous Nernst effect (ANE) \cite{Reichlova2019ImagingMn3Sn}, MOKE \cite{Higo2018LargeMetal}, and large tunneling magnetoresistance in MTJs \cite{Qin2023Room-temperatureJunction, Chen2023Octupole-drivenJunction}. Time-resolved studies have further revealed ultrafast magnetic phenomena in these materials, including coherent spin dynamics \cite{Miwa2021GiantSemimetal}, ultrafast magnetic quenching \cite{Miwa2021GiantSemimetal, Zhao2021LargeMn3Sn}, spin-current generation \cite{Song2025UltrafastAntiferromagnets}, and spin–torque-driven magnetic switching \cite{Lee2025Spin-torque-drivenMn3Sn}, highlighting the intrinsically fast magnetic response of nc-AFs.

A second, considerably less explored family consists of cubic compounds Mn$_3$Y (Y = Ir, Pt, Rh) and cubic antiperovskites Mn$_3$ZN (Z = Ga, Ni, Sn), which exhibit a triangular spin configuration within the (111) plane. Owing to their symmetry, these materials can also display AHE \cite{Boldrin2019AnomalousFilms, Liu2018ElectricalTemperature}, ANE \cite{Johnson2022IdentifyingMicroscopy, Beckert2023AnomalousFilms}, and MOKE \cite{Johnson2023Room-temperatureSignatures}. In addition, their strong magnetostructural coupling gives rise to a range of functionalities, including piezomagnetic \cite{Lukashev2008TheoryAntiperovskites, Boldrin2018GiantMn3NiN}, piezospintronic \cite{Guo2020GiantMetal, Yan2019AFields}, and magnetocaloric effects \cite{Zemen2017FrustratedTheory, Shi2016BaromagneticAnalysis, Matsunami2015GiantMn3GaN}. Despite these attractive properties, their ultrafast magnetic dynamics remain largely unexplored. In our previous work \cite{KimakArXiv:2601.07753}, we demonstrated ultrafast, non-thermal control of spin order in Mn$_3$NiN and Mn$_3$GaN driven purely by the polarization orientation of linearly polarized femtosecond laser pulses. Symmetry analysis and microscopic modeling indicated that optically induced torques alone could not fully account for the observed dynamics, and we proposed the formation of a transient spin spiral as a possible excitation mechanism \cite{KimakArXiv:2601.07753}.

In the present work, we investigate a complementary experimental regime, namely laser-induced magnetic dynamics that is independent of the polarization of the excitation pulses in the antiperovskites Mn$_3$NiN and Mn$_3$GaN. Using a pump–probe technique, in both materials we detect pronounced probe-polarization rotation dynamics, which is strongly dependent on the sample tilt relative to the direction of both the applied magnetic field and the probe beam. Owing to the different magnetic phases of Mn$_3$NiN and Mn$_3$GaN, their measured responses to the applied magnetic field strongly differ. In Mn$_3$NiN, we identify a field-dependent MOKE-like signal in the measured probe polarization rotation, which enables us to separate dynamics of magnetic order quenching from other contributions to the detected signal. In Mn$_3$GaN, whose non-collinear magnetic phase allows only magneto-optical effects quadratic in \textbf{M}, the magnetic signals do not depend on the applied magnetic field. Our theoretical modeling demonstrates that a correct interpretation of the measured dynamics requires accounting for the population of magnetic domains and its modification by the applied magnetic field. Within this framework, we successfully reproduced the observed dependencies on magnetic field, sample tilt and input probe polarization. We also revealed that in Mn$_3$NiN the quenching dynamics slows down significantly at elevated temperatures, in close analogy to conventional metallic FMs \cite{Koopmans2010ExplainingDemagnetization}, but in a contrast to non-collinear Heusler antiferromagnet Mn$_3$Sn \cite{Zhao2021LargeMn3Sn}.

\section{\label{sec:level2} Experimental details and materials}

\begin{figure*}
\includegraphics[width=0.9\textwidth]{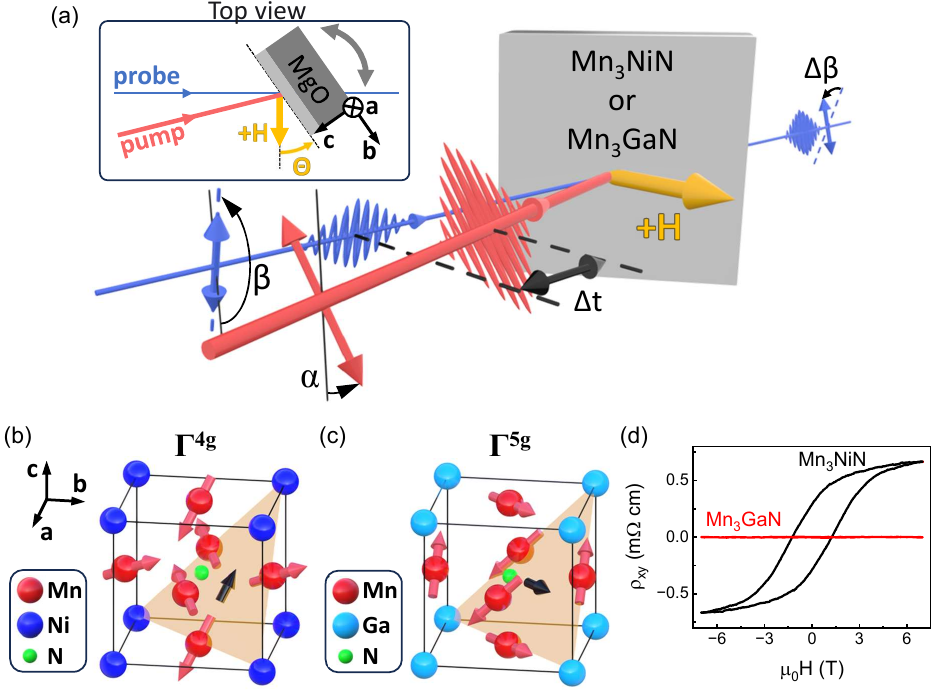}
\caption{\label{fig_1} (a) Schematics of the experimental pump-probe setup used for inducing and detecting magnetic dynamics in samples of Mn$_3$NiN or Mn$_3$GaN. Orientations of linear polarization of pump and probe pulses are described by angles $\alpha$ and $\beta$, respectively. The pump-induced change of polarization rotation $\Delta \beta$ of the transmitted probe pulses is measured as a function of the time delay $\Delta t$ between the pump and probe pulses. Inset: Simplified top view of the setup. The samples can be tilted around the crystallographic \textbf{a} axis by an angle $\Theta$ between the magnetic field \textbf{H} and the plane of the samples. Direction of \textbf{H} is always perpendicular to the probe beam. (b), (c) The crystallographic and magnetic structures of $\Gamma^{4g}$ phase in Mn$_3$NiN and $\Gamma^{5g}$ phase in Mn$_3$GaN, respectively. The three Mn spins (red arrows) are aligned within the (111) plane. The black arrows (not to scale) represent the small piezo-induced magnetic moment $\mathbf{M_P}$. (d) Electrical AHE measurement in Mn$_3$NiN and Mn$_3$GaN with the OOP direction of \textbf{H} at temperature 100\,K.}
\end{figure*}

We investigated time-resolved dynamics of the nc-AFs Mn$_3$NiN and Mn$_3$GaN using a pump–probe technique schematically shown in Fig.~\ref{fig_1}(a). The experiments were performed using a Ti:sapphire oscillator delivering $\approx150\,$fs laser pulses at a repetition rate of 80\,MHz. A portion of the oscillator output was used as the pump beam and focused onto the sample to a spot diameter of $\approx15\,\mu$m (full width at half maximum, FWHM), with a fluence of $\approx6\,\text{mJ/cm}^2$. The remaining part of the output was directed through an optical parametric oscillator to generate the probe beam, which was focused to a smaller spot with a FWHM of $\approx4\,\mu$m and a fluence of $\approx0.2\,\text{mJ/cm}^2$. The pump and probe wavelengths were set to $820\,$nm and $532\,$nm, respectively, or to $800\,$nm and $400\,$nm, respectively. For wavelengths of 800\,nm and 400\,nm for pump and probe beams, respectively, the repetition rate was reduced to 8\,MHz using a pulse picker. The orientations of the linear polarizations of the pump and probe beams, denoted by angles $\alpha$ and $\beta$, respectively, were controlled using $\lambda/2$ waveplates. The pump-induced change of the polarization rotation of the transmitted probe beam, $\Delta\beta$, was measured as a function of the pump–probe time delay, $\Delta t$, by detecting the difference between signals from two photodetectors in an optical bridge configuration (see Appendix B in \cite{Surynek2020InvestigationExperiment} for details). Simultaneously, we recorded the sum of the detector signals, which reflects changes in the probe intensity and is referred to as transient transmission, $\Delta \textsf{T}/\textsf{T}$, arising from pump-induced modifications of the sample transmission $\textsf{T}$.
 
The angle between the pump and probe beams was fixed at $\approx15^\circ$. The probe beam propagated perpendicularly to the direction of the applied magnetic field, \textbf{H}, with a maximum strength of $\mu_0 H = \pm530\,$mT. The samples were mounted in a cryostat allowing control of the base temperature $T$ over a wide range from 25\,K to 800\,K. In addition, the sample holder enabled tilting of the samples by an angle $\Theta$ about the axis perpendicular to both the applied magnetic field and the propagation directions of the beams, as illustrated in the inset of Fig.~\ref{fig_1}(a). 

A thin film of (001)-oriented Mn$_3$NiN with a thickness of 13\,nm was grown on a (001)-oriented MgO substrate by the pulsed laser deposition technique with no annealing. The Néel temperature for this sample, $T_N \approx 246\,$K was determined from magnetic characterization measurements performed using a SQUID magnetometer. X-ray diffraction (XRD) and Reciprocal space mapping measurements yielded lattice parameters of $c = 3.908$\,\r{A} and $a = 3.940$\,\r{A}, in accord with a slight tensile strain (see Supplementary Note 1 in \cite{KimakArXiv:2601.07753}). The resulting tetragonality of the unit cell, $c/a = 0.992$, is expected to induce a small net magnetic moment in the case of $\Gamma^{4g}$ phase [see Fig.~\ref{fig_1}(b)], via the piezomagnetic effect \cite{Boldrin2018GiantMn3NiN}. This moment, $\mathbf{M_P}$, points opposite to the spin of Mn atom lying in the (\textbf{a},\textbf{b}) plane, i.e. along $\left[-1, 1, 2\right]$ [see Fig.~\ref{fig_1}(b)]. $\mathbf{M_P}$ is coupled to the antiferromagnetic ordering and allows control of the domain structure with an applied magnetic field \cite{Boldrin2019TheFilms, Boldrin2019AnomalousFilms, Johnson2022IdentifyingMicroscopy}. However, $\mathbf{M_P}$ should not be confused with the magnetic octupole moment, present even in unstrained $\Gamma^{4g}$ phase, which plays an essential role as the magnetic order parameter for the AHE and related effects and which can be associated with an effective magnetic dipole moment along the $\left[1, -1, 1\right]$ direction \cite{Suzuki2017ClusterAntiferromagnets} in the coordinate system shown in Fig.~\ref{fig_1}(b). 
A (001)-oriented Mn$_3$GaN thin film with a thickness of 40\,nm was grown on a (001)-oriented MgO substrate by the physical vapor deposition technique with no annealing. For this sample, $T_N \approx 320\,$K was measured using a vibrating-sample magnetometer. XRD measurements confirmed a crystalline quality comparable to that of the Mn$_3$NiN film (see Supplementary Note 1 in \cite{KimakArXiv:2601.07753}). We expect a tensile strain for the Mn$_3$GaN sample as well, since the reported lattice constant of perfectly cubic Mn$_3$GaN is $\approx 3.86$\,\r{A} \cite{Ishino2018PreparationCompositions, Abbas2025ExperimentalMgO001}, while for MgO the lattice constant is $4.21$\,\r{A}. In the case of $\Gamma^{5g}$ phase [see Fig.~\ref{fig_1}(c)], the induced $\mathbf{M_P}$ points along the direction of the spin of Mn atom lying in the (\textbf{a},\textbf{b}) plane \cite{Zemen2017PiezomagnetismNitrides}, i.e. along $\left[1, 1, 0\right]$. Directions of $\mathbf{M_P}$ for both compounds and all 8 magnetic domain variants are summarized in Tables~\ref{table1} and \ref{table2} in Appendix~\ref{app:2}. Both samples were glued to the sample holder with the same crystallographic orientation in such a way that for $\Theta = 0$°, the crystallographic axes \textbf{c} and \textbf{b} are opposite of the probe beam and along +\textbf{H}, respectively [see the inset of Fig.~\ref{fig_1}(a)]. 

Since both compounds can adopt either the $\Gamma^{4g}$ or $\Gamma^{5g}$ magnetic phases depending on growth conditions and external parameters \cite{Boldrin2019AnomalousFilms, Boldrin2019TheFilms}, we characterized these films using magneto-transport measurements of AHE. The measurements were performed at a temperature of $T = 100\,$K with the magnetic field \textbf{H} applied along the out-of-plane (OOP) direction. In Mn$_3$NiN, the transverse resistivity $\rho_{xy}$ exhibits a clear hysteresis as a function of \textbf{H} after subtraction of the linear background, whereas no detectable signal is observed in Mn$_3$GaN [see Fig.~\ref{fig_1}(d)]. This behavior indicates that Mn$_3$NiN is in the $\Gamma^{4g}$ phase, which allows for non-zero Berry curvature
and hence for effects linear in \textbf{M}, while Mn$_3$GaN adopts the $\Gamma^{5g}$ phase, with zero Berry curvature \cite{Gurung2019AnomalousAntiperovskites}. A more comprehensive structural and magnetic characterization of the films is provided in the Supplementary Note 1 of our previous work \cite{KimakArXiv:2601.07753}.

In the used transmission geometry for the time-resolved MO measurements, pump-induced changes of a material give rise to a polarization rotation of probe pulses via various optical and MO effects. Pump-induced magnetic changes in both materials were detected through the Voigt effect (also known as the Cotton-Mouton effect or quadratic MOKE), as in our previous work on Mn$_3$NiN and Mn$_3$GaN \cite{KimakArXiv:2601.07753}, ferromagnetic GaMnAs \cite{Tesarova2014SystematicGaMnAs}, in nc-AF Mn$_3$Sn \cite{Zhao2021LargeMn3Sn}, and also in collinear AFs CuMnAs \cite{Saidl2017OpticalAntiferromagnet} and CoO \cite{Zheng2018Magneto-opticalFilms}. As described in detail in Ref.~\cite{Tesarova2014SystematicGaMnAs}, the dependence of the Voigt-signals is even in \textbf{\textit{M}}, which makes it particularly useful for studying AFs, even collinear ones \cite{Saidl2017OpticalAntiferromagnet, Zheng2018Magneto-opticalFilms}. Dependence on the light angle of incidence (AOI) is usually weak for the Voigt effect. In addition, in Mn$_3$NiN, we detected pump-induced changes of magnetic ordering also through MOKE-like signals in transmission geometry, which are odd in \textit{\textbf{M}} and are commonly used to study FMs and other spin-polarized magnetic materials.
Moreover, pump pulses induce also non-magnetic material changes, such as heating. The corresponding polarization rotation can be understood in terms of a modification of the isotropic refractive index (in cubic materials). At non-normal incidence of the probe beam, this leads to different indices of refraction for $s$~($\beta=0$°) and $p$ ($\beta=90$°) polarization orientations of the beam.

In this work, we focused solely on the pump-induced dynamics that is independent of the polarization of the pump pulses. As helicity dependent effects are not apparent in Mn$_3$NiN and Mn$_3$GaN (see Figs.~1b and 3c in \cite{KimakArXiv:2601.07753}), the polarization-independent dynamics can be equivalently measured with excitation by left- or right-circularly polarized pump pulses, or by averaging signals obtained with two orthogonal orientations of pump linear polarization $\alpha$. An example of the as-measured data with $\alpha=0$° and $\alpha=90$°, as well as the corresponding pump-polarization-dependent (PPD) and pump-polarization-independent (PPI) signal parts are shown in Fig.~\ref{fig_6} in Appendix \ref{app:1}.

\section{\label{sec:level3} Results and discussion}
\subsection{\label{sec:level3_1} Pump-induced dynamics of polarization rotation and its origins}

\begin{figure}
\includegraphics[width=0.45\textwidth]{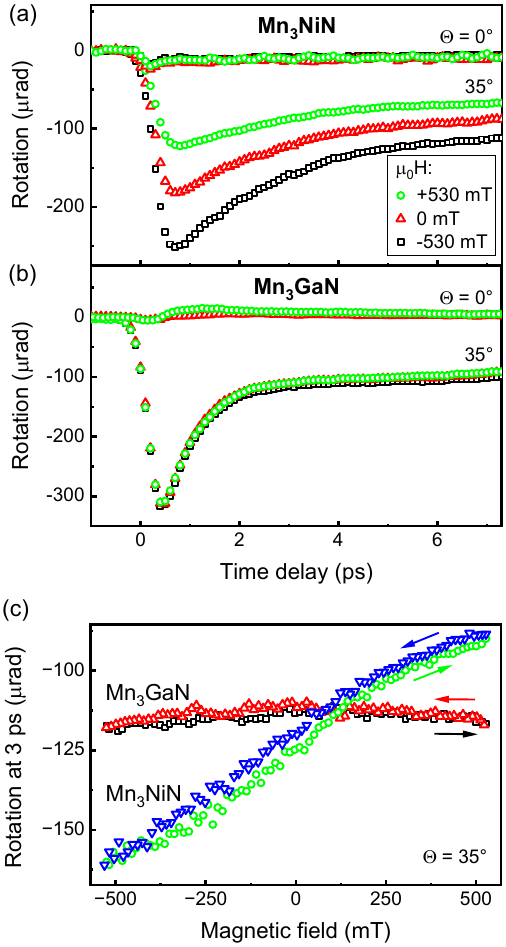}
\caption{\label{fig_2} Dynamics of pump-polarization-independent part of the probe polarization rotation in samples of (a) Mn$_3$NiN and (b) Mn$_3$GaN for various applied magnetic fields and sample tilts $\Theta = 0$° and $35$°. (c) Magnetic field dependence of the rotation at a time delay $\Delta t = 3\,$ps at $35^\circ$ tilt. Probe with a wavelength 532\,nm was polarized along $\beta = 135^\circ$, samples temperature 100\,K.}
\end{figure}

In Fig.~\ref{fig_2}(a) we show representative data of pump-induced change of the probe polarization rotation in Mn$_3$NiN measured for sample tilts $\Theta = 0$° and 35° using probe polarization $\beta = 135$°. For normal probe incidence ($\Theta = 0$°), when the applied magnetic field is purely in-plane (IP), the rotation signals are nearly absent. In contrast, tilting the sample to $\Theta = 35$° gives rise to pronounced dynamics, which depends significantly on the applied magnetic field. Such dynamics are characteristic of metallic magnets, where femtosecond optical excitation rapidly heats the electronic subsystem, which subsequently transfers energy to the phonon and spin subsystems, leading to a quenching of magnetic order \cite{Kirilyuk2010UltrafastOrder}. Simultaneously, lattice heating modifies the refractive index, giving rise to the probe polarization rotation of purely non-magnetic origin \cite{Sun1994Femtosecond-tunableGold, DelFatti1998NonequilibriumFilms}.

To elucidate the roles of sample tilt and magnetic field on the measured rotation signals, we performed the same measurements in Mn$_3$GaN sample, which adopts a different magnetic structure [see Figs.~\ref{fig_1}(b) and \ref{fig_1}(c)]. As shown in Fig.~\ref{fig_2}(b), a strong dependence on sample tilt is again observed, whereas the magnetic-field dependence is absent in Mn$_3$GaN. The field dependence of the rotation at $\Theta=35$° for a fixed delay of 3\,ps is summarized in Fig.~\ref{fig_2}(c). Over the entire field range of $\pm 530\,$mT, the rotation in Mn$_3$GaN remains virtually field-independent, while in Mn$_3$NiN, it exhibits a distinct component that is odd in the magnetic field. This different behavior is consistent with the different magnetic structures in these two materials: the $\Gamma^{4g}$ phase in Mn$_3$NiN, unlike the $\Gamma^{5g}$ phase in Mn$_3$GaN, allows the existence of magneto-optical effects linear in the magnetic order parameter, such as AHE and MOKE [see Fig.~\ref{fig_1}(d)].

As we will show below in our modeling, the strong dependence of polarization rotation dynamics on $\Theta$ results from (i) dependencies of MOKE-like and Voigt signals, and signal due to refractive index change on AOI of the probe beam and (ii) dependence of domain population on the angle between \textbf{H} and the sample plane. In Mn$_3$NiN at $\Theta = 0$°, purely IP magnetic field populates four domains equally more than the rest of the domains. However, as at this sample tilt the probe is incident normally on the sample, the contributions of these four domains cancel exactly in the total MOKE-like response and nearly cancel in the Voigt response; the same signal compensation occurs within the remaining four domains. Tilting the sample to $\Theta \neq 0$° introduces an OOP component of \textbf{H}, lifting the degeneracy of domain population. At the same time, the non-normal probe incidence modifies the individual domain contributions to the total MOKE-like and Voigt signals, resulting in a net field-dependent MO response. In Mn$_3$GaN at $\Theta = 0$°, the Voigt effect is almost exactly compensated for the same reason as in Mn$_3$NiN, while the MOKE-like signals are forbidden due to the material magnetic phase symmetry. However, in contrast to Mn$_3$NiN, OOP component of the magnetic field at $\Theta \neq 0$ does not lift the degeneracy of domain population in Mn$_3$GaN, because of different directions of $\mathbf{M_P}$ w.r.t. \textbf{H} [see Figs.~\ref{fig_1}(a)-(c) and Tables~\ref{table1} and \ref{table2}]. This results in an increase of polarization rotation signal due to the Voigt effect, which is, however, independent of the applied field. Lastly, the dependence of signals on the sample tilt in both compounds arises also from the (non-magnetic) changes of the refractive index, whose contribution to polarization rotation in cubic crystals is finite only for non-normal probe incidence.

\subsection{\label{sec:level3_2} Modeling of optical and MO responses}

\begin{figure}
\includegraphics[width=0.45\textwidth]{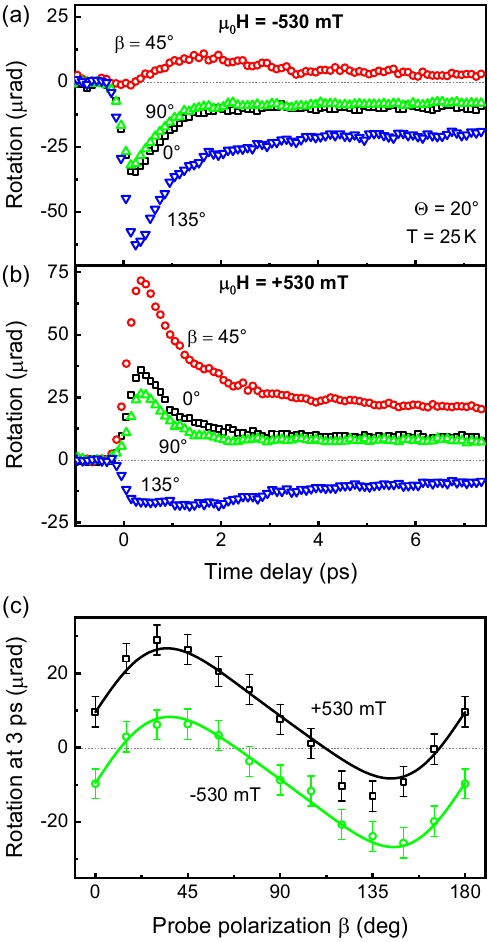}
\caption{\label{fig_3} Dynamics of rotation in Mn$_3$NiN measured for various orientations of input probe polarization $\beta$ for magnetic fields (a) --530\,mT and (b) +530\,mT at temperature 25\,K, sample tilt $20^\circ$ and probe wavelength 400\,nm. (c) $\beta$-dependence extracted from (a) and (b) at a time delay 3\,ps (symbols). Curves are fits by the model described in the text and Fig.~\ref{fig_4}.}
\end{figure}

\begin{figure*}
\includegraphics[width=0.9\textwidth]{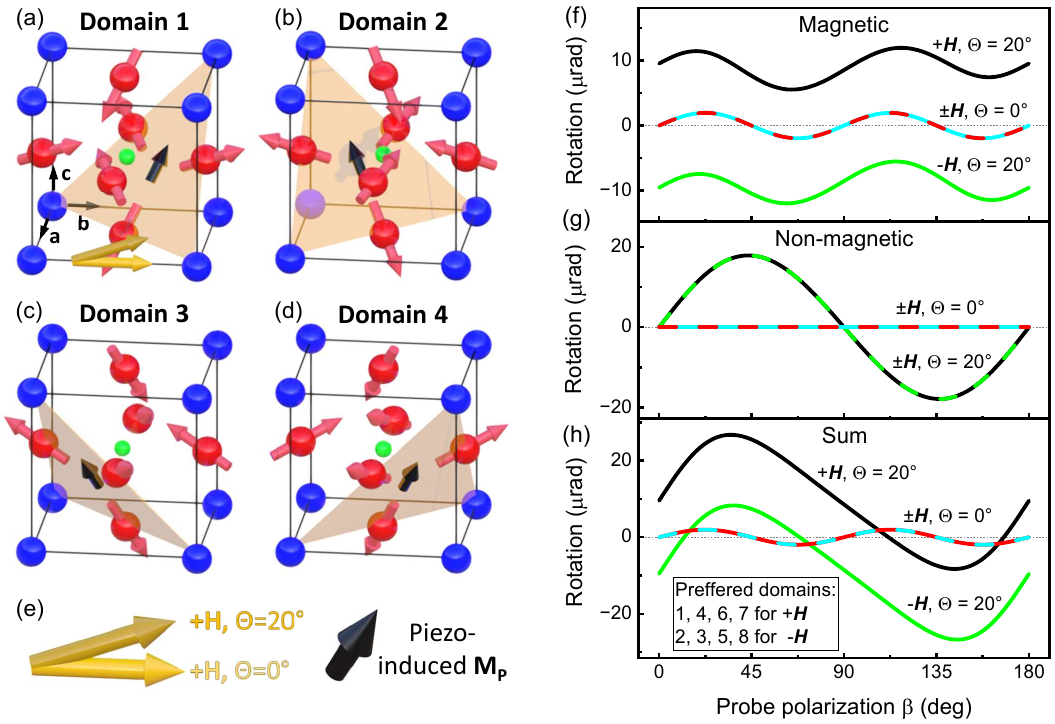}
\caption{\label{fig_4} (a)-(d) 4 of the 8 variants of the $\Gamma^{4g}$ phase. The remaining domains are shown in Fig.~\ref{fig_7}. (e) Legend for (a), where  directions of magnetic field +\textbf{H} for sample tilts 20° and 0° are depicted with a dark-yellow and a light-yellow arrow, respectively. Black arrow represents the piezo-induced magnetic moment $\mathbf{M_P}$, through which the magnetic field can prefer population of certain domains. (f)-(h) Computed $\beta$-dependence of probe polarization rotation reflecting (f) pump-induced magnetic quenching by 50\,\%, (g) pump-induced increase of the non-magnetic permittivity component $\epsilon^{\left(0\right)}$ by 0.2\,\%, and (h) their sum for 4 domain configurations, as described in detail in Appendix~\ref{app:2}. Curves in (h) for $\pm$\textbf{H} at tilt 20° are replicated in Fig.~\ref{fig_3}(c).}
\end{figure*}

Since distinct optical and MO effects exhibit different dependencies on the orientation of the incident probe polarization $\beta$, measuring the $\beta$-dependence and comparing it with a theoretical model provides a powerful means of disentangling the contributions to the observed signals \cite{Schmoranzerova2022GiantExperiments, Wohlrath2025QuadraticAnisotropies}. Because our model is formulated for zero temperature, we performed the $\beta$-dependence measurements at our lowest temperature of $25\,K$. Figures~\ref{fig_3}(a) and \ref{fig_3}(b) show, for magnetic fields of +530\,mT and --530\,mT, respectively, the rotation dynamics in Mn$_3$NiN for several selected values of $\beta$, measured at a sample tilt of $\Theta=20$°. The corresponding $\beta$-dependence extracted at a fixed delay of 3\,ps for all measured values of $\beta$ is shown by symbols in Fig.~\ref{fig_3}(c).

We modeled the optical and MO response using Yeh’s formalism \cite{Visnovsky2006OpticsNanostructures}, which accounts for the full multilayer (film + substrate) response in the relevant experimental geometry and therefore includes all optical and MO effects simultaneously. For the $\Gamma^{4g}$ phase in domain~1 [Fig.~\ref{fig_4}(a)], the permittivity tensor has the form
\begin{equation} 
\label{eq_permittivity} 
\epsilon = 
\begin{pmatrix} 
\epsilon^{\left(0\right)} +Q & 0 & 0 \\ 0 & \epsilon^{\left(0\right)} & +K \\ 0 & -K & \epsilon^{\left(0\right)} 
\end{pmatrix}
\end{equation}
in the coordinate system with $(y,z)$ plane lying in the plane of the spins, i.e. with $x\parallel\left[ 1, -1, 1 \right]$, $y\parallel\left[2,1,-1 \right]$ and $z\parallel\left[0,1,1 \right]$. Both parameters $K$ and $Q$ quantify the amplitude of the MO signals corresponding to changes of magnetic ordering, however, MOKE-like signals are present due to the parameter $K$, which is odd in \textit{\textbf{M}}. The Voigt-rotation arises from both $Q$, which is even in M, and $K$. The non-magnetic parameter $\epsilon^{\left(0\right)}$ is related to the complex index of refraction of the material. The tensors for the remaining seven domains [Figs.~\ref{fig_4}(b)–\ref{fig_4}(d) and Fig.~\ref{fig_7}] were generated by rotating Eq.~(\ref{eq_permittivity}) about the \textbf{c} axis and, for domains~5–8, inverting the sign of $K$. Pump-induced magnetic quenching was simulated by reducing the magnetic parameters $Q$ and $K$, whereas pump-induced change of refractive index due to heating was captured by modifying the non-magnetic component $\epsilon^{\left(0\right)}$. The effect of the applied magnetic field was incorporated through a redistribution of domain populations prior to excitation by a pump pulse, i.e. by forming a weighted sum of the total optical and MO responses over all domains. This is possible due to a presence of piezo-induced $\mathbf{M_P}$, which is coupled to the antiferromagnetic ordering \cite{Boldrin2019TheFilms, Boldrin2019AnomalousFilms, Johnson2022IdentifyingMicroscopy}. In total, the optical and MO response was parameterized by (i) the magnitude of pump-induced magnetic quenching, (ii) the pump-induced change of the refractive index, and (iii) the field-induced change in domain populations for opposite magnetic-field polarities. A detailed description of the model and the fitting procedure is provided in Appendix~\ref{app:2}.

The modeled $\beta$-polarization dependence of the total optical and MO response [solid curves in Fig.~\ref{fig_3}(c)] reproduces the measured data for sample tilt $\Theta=20$° very well. The individual magnetic and non-magnetic contributions corresponding to the best fit are shown with black and green curves in Figs.~\ref{fig_4}(f) and \ref{fig_4}(g), respectively, and their sum in Fig.~\ref{fig_4}(h). As expected, magnetic quenching leads to a deviation from a sinusoidal waveform of dependence on $\beta$, which comes mainly from the contribution of the (non-magnetic) refractive index change. In this geometry, the +\textbf{H} field increases the population of domains~1 and~4 and, to a smaller extent, also domains~6 and~7. Correspondingly, the populations of domains~2 and~3 decrease, as do domains~5 and~8, which experience the strongest reduction. This domain redistribution follows from the different angles between $\mathbf{M_P}$ and +\textbf{H} for the individual domains (see Table~\ref{table1} in Appendix~\ref{app:2}). As a consequence, the domain population is imbalanced and the weighted sum of individual domains contributions to total MOKE-like and Voigt signals [see Figs.~\ref{fig_9} (c), (d)] results in a sizable net polarization rotation signals [Fig.~\ref{fig_4}(f)]. Analogically for \textbf{--H} field. This mechanism also explains the observed dependence of the measured MO signal on the magnitude of the applied field [Fig.~\ref{fig_2}(c)].

From the fit of the experimental data we obtain a quenching magnitude of $(50\pm20)\%$. As a part of the modeling, assuming the same magnetic quenching, we also simulated the configuration $\Theta = 0$° [red and blue curves in Figs.~\ref{fig_4}(f)–\ref{fig_4}(h)], corresponding to the very small measured signals in Fig.~\ref{fig_2}(a). In this geometry, the IP +\textbf{H} field increases the population of domains~1, 4, 6 and~7 to the same extent, and, correspondingly, the domains~2, 3, 5 and~8 experience the same reduction (see Table~\ref{table1} in Appendix~\ref{app:2}). As a consequence, the domain populations is much more balanced compared to $\Theta = 20$°. Moreover, MOKE-like and Voigt signals from individual domains are more degenerate at normal probe incidence [see Figs.~\ref{fig_9}(a), \ref{fig_9}(b)] than for $\Theta=20$° [see Figs.~\ref{fig_9}(c), \ref{fig_9}(d)]. The weighted sum of individual domains contributions to the total MOKE-like signal therefore exactly cancels out. For the Voigt effect, the corresponding weighted sum yields only a small polarization rotation, below 3\,$\mu$rad [see Fig.~\ref{fig_4}(f)]. Analogically for \textbf{--H} field. Thus, it is the combination of normal probe incidence and a purely IP magnetic field that leads to the almost complete disappearance of MO signals at $\Theta = 0$°. 

The same modeling framework can be used to rationalize the signals in Mn$_3$GaN by assuming vanishing odd-in-\textit{\textbf{M}} parameter $K$, consistent with the $\Gamma^{5g}$ magnetic structure. The relative magnitudes of the parameters of the permittivity tensor differ for Mn$_3$GaN, but the Voigt effect arising solely from the parameter $Q$ is assumed to have the same $\beta$-symmetry in $\Gamma^{4g}$ and $\Gamma^{5g}$ phases, so a qualitative description can be obtained by setting $K=0$. Similarly to the relation between magnetic phases in Figs.~\ref{fig_1}(b) and \ref{fig_1}(c), the eight Mn$_3$GaN domains (see Fig.~\ref{fig_8}) can be mapped from the Mn$_3$NiN domains by a 90° spin rotation in the spin plane. For $\Theta = 20$° [and similarly for $\Theta = 35$°, as in Fig.~\ref{fig_2}(b)], the applied +\textbf{H} field increases the population of domains~1, 2, 7, and~8 by the same amount and, correspondingly, reduces the populations of the remaining domains uniformly (see Table~\ref{table2} in Appendix~\ref{app:2}). This behavior is in contrast with Mn$_3$NiN, where for $\Theta \neq 0$ the +\textbf{H} field leads to an asymmetric redistribution, with two domains being enhanced more strongly than the other favored pair (see Table~\ref{table1} in Appendix~\ref{app:2}); an analogous redistribution occurs for \textbf{--H}. This difference between Mn$_3$NiN and Mn$_3$GaN arises from the different orientations of the piezo-induced $\mathbf{M_P}$ and, consequently, from the different angles between $\mathbf{M_P}$ and $\pm$\textbf{H} in the two compounds (see Tables~\ref{table1} and \ref{table2} in Appendix~\ref{app:2}). Moreover, because the Voigt effect is even w.r.t the magnetic moments directions, domain variants with opposite moment directions have equal contributions to the polarization rotation due to the Voigt effect. In Mn$_3$GaN, the total Voigt contribution from domains~1, 2, 7, and~8 is therefore equal to that from the remaining domains. As a result, changes of domain populations induced by changing amplitude of \textbf{H} do not modify the polarization rotation due to the Voigt effect, yielding a finite but field-independent signal (see black and green curves in Fig.~\ref{fig_11} in Appendix~\ref{app:2}), consistent with the behavior observed in Fig.~\ref{fig_2}(b). However, the (non-magnetic) refractive index change also contributes a field-independent rotation, thus separating its contribution from the Voigt response in the measured dynamics would require a different experimental approach.

For $\Theta = 0$°, the same domains (1, 2, 7 and~8) in Mn$_3$GaN remain equally populated for +\textbf{H}. However, under normal probe incidence their individual Voigt-induced polarization rotations are degenerate [see Fig.~\ref{fig_9}(b)], and the same holds for the remaining domains. As a result, the weighted sum of all domain contributions yields only a small net rotation, on the order of $\mu$rad (see the red and blue curves in Fig.~\ref{fig_11} in Appendix~\ref{app:2}). This behavior is analogous to Mn$_3$NiN and is consistent with the experimentally observed response in Fig.~\ref{fig_2}(b).

\subsection{\label{sec:level3_3} Single- and two-step quenching dynamics}

\begin{figure}
\includegraphics[width=0.45\textwidth]{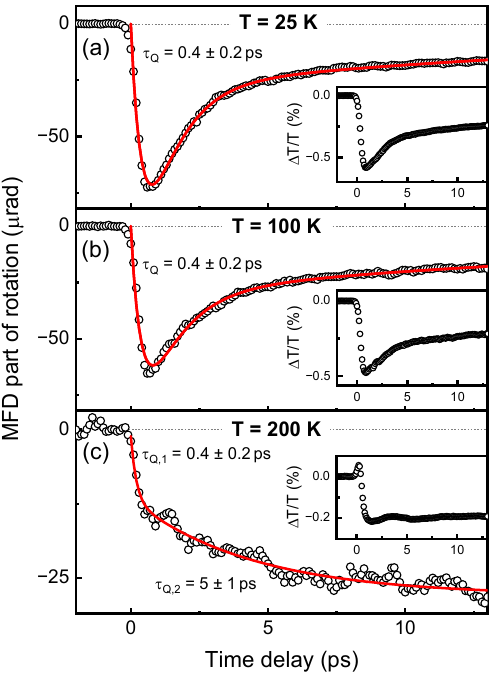}
\caption{\label{fig_5} Dynamics of magnetic-field-dependent part of the rotation signal in Mn$_3$NiN at various temperatures (symbols). At temperatures 25\,K (a) and 100\,K (b), the signals show a single-step quenching, typical for type-I demagnetization, while at 200\,K (c), the signal shows a two-step quenching, typical for type-II demagnetization. Curves: Fits by Eqs.~(\ref{eq_fit}) and (\ref{eq_fit2}) with $A=160\pm20 \,\mu$rad, $B=130\pm20 \,\mu$rad, $\tau_Q = 0.4 \pm 0.2$\,ps, $\tau_1 = 1.1 \pm 0.5$\,ps, $\tau_2 = 23 \pm 5$\,ps for 25\,K; $A=120 \pm 15 \,\mu$rad, $B=90 \pm 15 \,\mu$rad, $\tau_Q = 0.4 \pm 0.2$\,ps, $\tau_1 = 1.3 \pm 0.4$\,ps, $\tau_2 = 25 \pm 5$\,ps for 100\,K; and $A_1 = 12 \pm 2 \,\mu$rad, $A_2 = 16 \pm 2 \,\mu$rad, $\tau_{Q,1} = 0.4 \pm 0.2$\,ps, $\tau_{Q,2} = 5 \pm 1 $\,ps  for 200\,K. Insets: Dynamics of the transient transmission at corresponding temperatures; x-scale is the same as in the main figures. Probe wavelength was 532\,nm, $\beta = 135$°, sample tilt $\Theta = 35$°.}
\end{figure}

To isolate the MOKE-like signals corresponding to dynamics of magnetic order in Mn$_3$NiN, we define the magnetic-field-dependent (MFD) component as $\text{MFD}~\equiv~\left( \Delta\beta_{-530mT}~\,~-~\Delta\beta_{+530mT} \right) /2$. This removes all the field-independent contributions, including the refractive index change and the optical transients, persisting for approximately 1.5\,ps in Mn$_3$NiN \cite{KimakArXiv:2601.07753}. The resulting MFD traces measured at three temperatures, namely 25\,K, 100\,K and 200\,K, shown with symbols in Figs.~\ref{fig_5}(a)–\ref{fig_5}(c), therefore isolate the dynamics associated solely with magnetic quenching over the full delay range. While the response at 25\,K and 100\,K exhibits a single-step quenching, the dynamics at 200\,K develops a distinct two-step character.

We fit the $\Delta t \geq 0 $ data at 25\,K and 100\,K using \cite{You2018RevealingSpectroscopies}
\begin{equation}
\label{eq_fit}
\Delta \beta \left( \Delta t\right) = A e^{-\Delta t/\tau_Q} - B e^{-\Delta t/\tau_{1}} + (B-A) e^{-\Delta t/\tau_{2}},
\end{equation}
where $\tau_Q$ denotes the characteristic quenching time, $\tau_{1}$ and $\tau_{2}$ describe the recovery dynamics, and $A$ and $B$ the corresponding amplitudes. In contrast, the 200\,K data are described more accurately by a two-step quenching function \cite{Gong2023UltrafastMagnet}
\begin{equation}
\label{eq_fit2}
\Delta \beta \left( \Delta t\right) = A_1 e^{-\Delta t/\tau_{Q,1}} + A_2 e^{-\Delta t/\tau_{Q,2}} - (A_1+A_2),
\end{equation}
with two quenching amplitudes $A_1$, $A_2$ and time constants $\tau_{Q,1}$, $\tau_{Q,2}$. The constant offset $-(A_1+A_2)$ reflects the absence of a resolvable recovery within the experimental time window. The prefactors in Eqs.~(\ref{eq_fit}) and (\ref{eq_fit2}) are chosen such that $\Delta \beta(\Delta t=0)=0$.

From the fits (solid curves in Fig.~\ref{fig_5}), we obtain $\tau_Q = 0.4 \pm 0.2$\,ps at 25\,K and 100\,K, comparable to values reported for ultrafast quenching/demagnetization in nc-AF Mn$_3$Sn \cite{Zhao2021LargeMn3Sn} and in metallic FMs such as Ni \cite{You2018RevealingSpectroscopies, Koopmans2010ExplainingDemagnetization, Roth2012TemperatureMechanism} and Co \cite{Koopmans2010ExplainingDemagnetization} at low temperatures, with type-I quenching. At 200\,K, where the MFD dynamics is reminiscent of type-II quenching observed in systems such as Tb$_{35}$Fe$_{65}$ \cite{Kim2009UltrafastFilm}, Ni and Co at high temperatures \cite{Roth2012TemperatureMechanism, Koopmans2010ExplainingDemagnetization}, we obtain a fast first quenching with $\tau_{Q,1} = 0.4 \pm 0.2$\,ps and a second, much slower quenching with $\tau_{Q,2} = 5 \pm 1$\,ps. This shows, in a contrast to the nearly temperature-independent quenching reported for Mn$_3$Sn \cite{Zhao2021LargeMn3Sn}, that in Mn$_3$NiN, the quenching exhibits a pronounced slowing down with increasing $T$, qualitatively similar to the behavior of metallic FMs \cite{Koopmans2010ExplainingDemagnetization}.

In metallic FMs, ultrafast quenching is commonly modeled using various combinations of two- \cite{Sultan2012Electron-Gd0001} and three-temperature models \cite{Koopmans2010ExplainingDemagnetization, Roth2012TemperatureMechanism, Shim2020RoleMultilayer, Pankratova2022Heat-conservingNickel, Beaurepaire1996UltrafastNickel}, microscopic Elliott–Yafet (EY) scattering \cite{Koopmans2010ExplainingDemagnetization, Atxitia2011UltrafastModel}, Landau–Lifshitz–Bloch equations \cite{Atxitia2011UltrafastModel, Chubykalo-Fesenko2020Landau-Lifshitz-BlochTransition}, and related frameworks \cite{Chen2025UltrafastProgress}. The slowing down at elevated temperatures can arise from several factors, including a reduced magnetic order parameter prior to excitation \cite{Koopmans2010ExplainingDemagnetization}, temperature dependence of susceptibility \cite{Chubykalo-Fesenko2020Landau-Lifshitz-BlochTransition}, weakened electron-spin coupling \cite{Roth2012TemperatureMechanism} or changes in phonon-mediated processes \cite{Koopmans2010ExplainingDemagnetization, Sultan2012Electron-Gd0001}. Notably, unlike the temperature-invariant transient reflectivity reported for Mn$_3$Sn \cite{Zhao2021LargeMn3Sn}, our transient transmission data (insets in Fig.~\ref{fig_5}) indicate a modified rate of energy transfer into the phonon subsystem in Mn$_3$NiN at higher temperatures. This suggests that changes in phonon-mediated processes, such as the aforementioned EY scattering, electron-lattice or spin-lattice couplings may contribute to the deceleration of quenching. At the same time, the microscopic nature of quenching in nc-AFs—whether dominated by a reduction of local moments, randomization of their orientations, or a combination of both—remains an open question, which is beyond the scope of this paper.

\section{\label{sec:level5} Conclusions}

Using the pump-probe technique, we have investigated pump-polarization-independent dynamics of MO signals in non-collinear antiferromagnetic antiperovskites Mn$_3$NiN and Mn$_3$GaN. In both materials, the measured probe polarization rotation dynamics show a strong dependence on the sample tilt with respect to the direction of the probe beam and the magnetic field. In Mn$_3$NiN, where the rotation dynamics additionally exhibits a magnetic-field dependence, we were able to separate contributions from pump-induced quenching of magnetic order and a pump-induced modification of the refractive index. By combining measurements for multiple input probe polarization orientations with modeling based on Yeh's formalism, we have satisfactorily explained the observed dependence on the input probe polarization, as well as on the sample orientation and on the magnitude and direction of the magnetic field, which is causing field-controlled redistribution of domain populations. Also in Mn$_3$GaN, which has a different magnetic phase, the applied magnetic field does modify the domain population. However, as the symmetry forbids the existence of MOKE-like effects, the MO signals are not influenced by such domain redistribution. Even though the magnetic quenching in Mn$_3$GaN probably contributes to the rotation dynamics, disentangling its magnetic and non-magnetic contributions is not straightforward, since the MOKE-like signals are absent. Finally, we showed that in Mn$_3$NiN the quenching dynamics changes from a fast, single-step (type-I), to a slower, two-step (type-II) quenching upon temperature increase. This behavior is similar to that observed in metallic FMs \cite{Koopmans2010ExplainingDemagnetization}, but distinct from the quenching dynamics reported for non-collinear antiferromagnetic Heusler compound Mn$_3$Sn \cite{Zhao2021LargeMn3Sn}.

\begin{acknowledgments}
This work was supported by TERAFIT project No. CZ.02.01.01$/$00$/$22\_008$/$0004594 funded by Ministry of Education Youth and Sports of the Czech Republic (MEYS CR), programme Johannes Amos Comenius (OP JAK), call Excellent Research, and by CzechNanoLab Research Infrastructure supported by MEYS CR (LM2023051), and by e-INFRA CZ (ID:90254) project of MEYS CR. F.J. is grateful for support from the Royal Commission of 1851 Research Fellowship. L.N. acknowledges the networking opportunities provided by the European COST Action No. CA23136 (CHIROMAG).
\end{acknowledgments}

\appendix

\section{\label{app:1} PPD and PPI parts of the as-measured data}

\begin{figure*}
\includegraphics[width=0.9\textwidth]{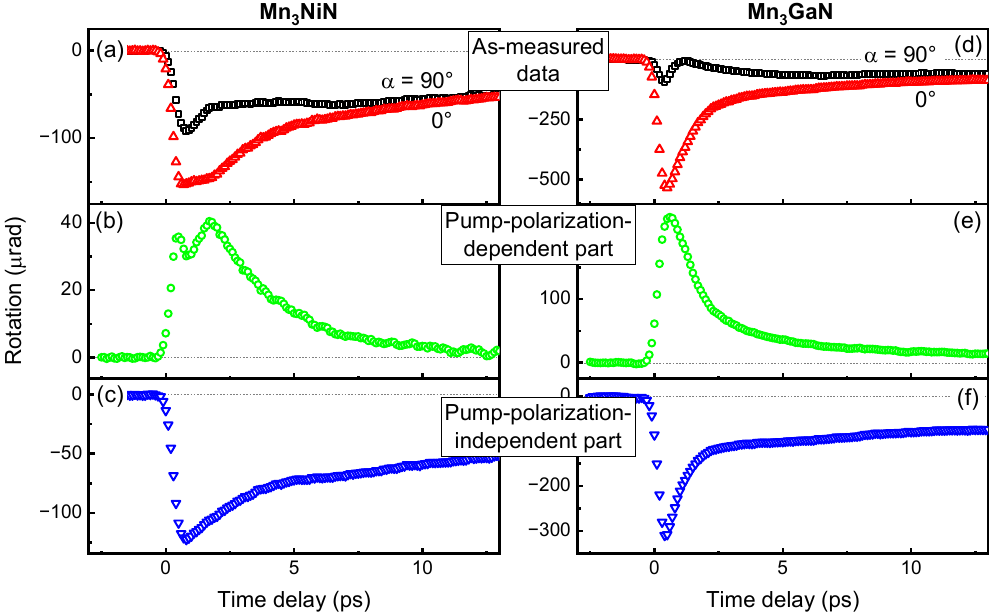}
\caption{\label{fig_6} (a) As-measured dynamics of rotation in Mn$_3$NiN for 2 orthogonal polarization orientations of pump pulses measured at temperature 100\,K with a magnetic field +530\,mT, sample tilt $35^\circ$, probe polarization $\beta = $135° and wavelength 532\,nm. (b) Pump-polarization-dependent part of the signal. (c) Pump-polarization-independent part of the signal. (d)-(f) Same as (a)-(c), but for Mn$_3$GaN.}
\end{figure*}

\begin{figure}
\includegraphics[width=0.45\textwidth]{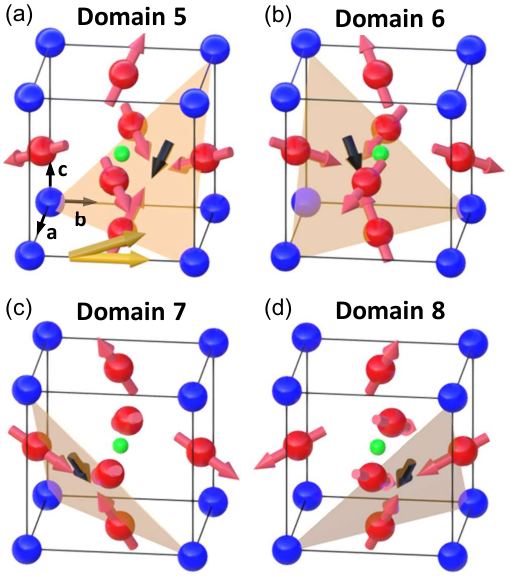}
\caption{\label{fig_7} (a)-(d) As in Fig.~\ref{fig_4} (a)-(d), showing the rest of the 8 variants of $\Gamma^{4g}$.}
\end{figure}

\begin{figure*}
\includegraphics[width=0.9\textwidth]{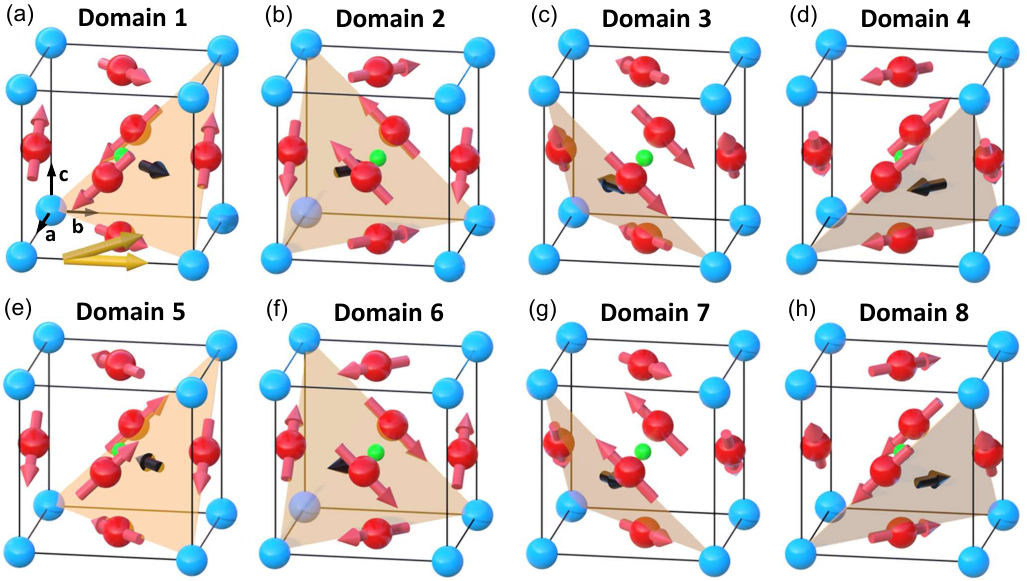}
\caption{\label{fig_8} 8 variants of $\Gamma^{5g}$ phase.}
\end{figure*}

\begin{figure*}
\includegraphics[width=0.9\textwidth]{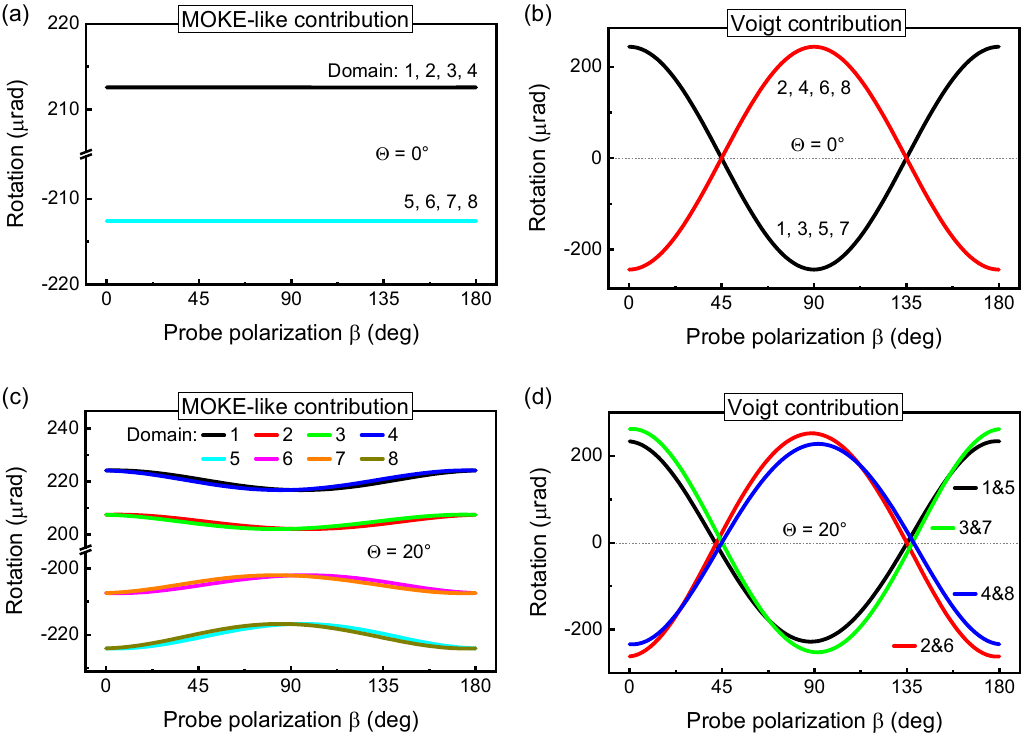}
\caption{\label{fig_9} Dependencies of (a) MOKE-like signals and (b) Voigt signals reflecting pump-induced 50\,\% magnetic quenching on input probe polarization orientation $\beta$ computed for all 8 variants of $\Gamma^{4g}$ domains and sample tilt 0°. (c), (d) Same as (a), (b), for sample tilt 20°. Weighted average of all the curves (with the weights defining the domain population described in Appendix~\ref{app:2_2}) corresponds to the curves displayed in Fig.~\ref{fig_4}(f).}
\end{figure*}

\begin{figure*}
\includegraphics[width=0.9\textwidth]{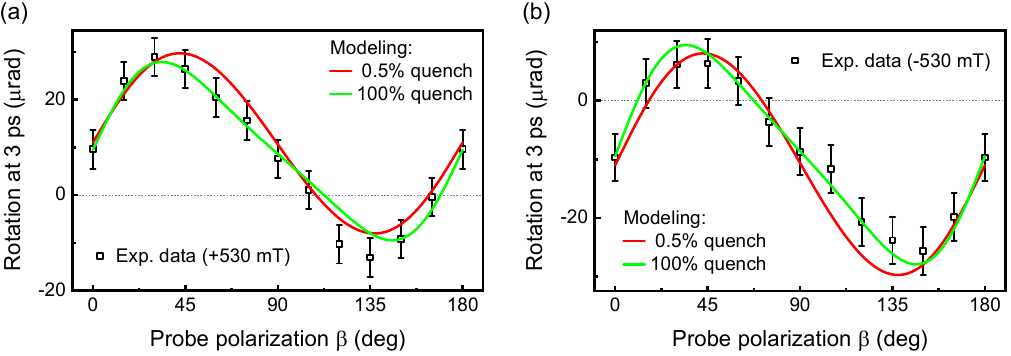}
\caption{\label{fig_10} Modeling the $\beta$-dependence of rotation with 0.5\% and 100\% pump-induced magnetic quenching (curves). Experimental data at 25\,K are shown with open squares. (a) Positive and (b) negative magnetic field.}
\end{figure*}

\begin{figure}
\includegraphics[width=0.45\textwidth]{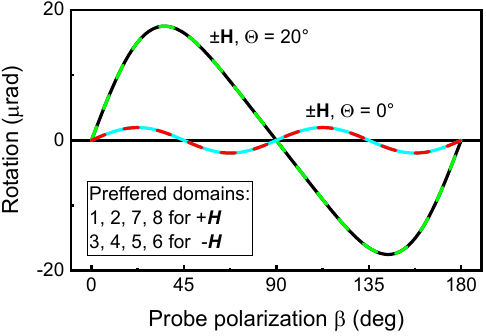}
\caption{\label{fig_11} As in Fig.~\ref{fig_4}(h), but for Mn$_3$GaN, with the $K$ parameter set to zero.}
\end{figure}

Magnetic dynamics in Mn$_3$NiN and Mn$_3$GaN exhibit a strong dependence on pump polarization orientation $\alpha$. Representative examples are shown in Figs.~\ref{fig_6}(a) and \ref{fig_6}(b), which display the rotation dynamics of Mn$_3$NiN and Mn$_3$GaN, respectively, for excitation with two orthogonal linear polarizations of the pump pulse. In our previous work \cite{KimakArXiv:2601.07753}, we analyzed in detail the pump-polarization-dependent (PPD) component of the signal, defined for probe polarization $\beta=135$° as $\left( \Delta \beta_{\alpha=90^\circ} - \Delta \beta_{\alpha=0^\circ} \right) / 2$ and shown in Figs.~\ref{fig_6}(b) and \ref{fig_6}(e). In the present work, we focused only on the pump-polarization-independent (PPI) component, defined as $\left( \Delta \beta_{\alpha=90^\circ} + \Delta \beta_{\alpha=0^\circ} \right) / 2$, which is shown in Figs.~\ref{fig_6}(c) and \ref{fig_6}(f). As noted earlier, this component is equivalent to as-measured data for excitation with a left- or right-circularly polarized pump pulse.

\section{\label{app:2} Modeling}
\subsection{\label{app:2_1} Permittivity tensor}

In order to calculate the parameters of the permittivity tensor, we first use non-collinear spin-polarized density functional theory (DFT) to find the electronic structure. The spin-orbit coupling is included and the local magnetic moments are constrained to $\Gamma^{4g}$ configuration (ground state). Subsequently, we employ linear response theory following the approach of \cite{Feng2015LargePt} and our previous work \cite{Johnson2023Room-temperatureSignatures,Zemen2023CollinearSpectroscopy}. We use the projector augmented wave method implemented in the VASP code \cite{Kresse1993AbMetals} with generalized gradient approximation (GGA) parametrized by Perdew–Burke–Ernzerhof. The plane-wave energy cutoff is 500~eV. The valence configurations of manganese, nickel and nitrogen are 3$p^6$3$d^6$4$s^1$, 3$p^6$3$d^9$4$s^1$, and 2$s^2$2$p^3$, respectively. We use a regular 20 × 20 × 20 reciprocal space mesh.

As has been done for MOKE spectra of Mn$_3$NiN \cite{Johnson2023Room-temperatureSignatures}, and earlier MOKE studies of some collinear antiferromagnets such as CuMnAs \cite{Veis2018BandSpectroscopy}, we modify the intra-atomic Coulomb interaction within GGA through the rotationally invariant approach to GGA+U as proposed by Dudarev et al. in \cite{Dudarev1998Electron-energy-lossStudy}. Our permittivity is simulated using $U$ = 0.7~eV, following the approach used in Ref.~\cite{Johnson2023Room-temperatureSignatures}. It lifts the unoccupied manganese $3d$-states further away from the Fermi level, resulting in a blue-shift in the MO responses and in better the agreement with the available measured data \cite{Johnson2023Room-temperatureSignatures}. 

\subsection{\label{app:2_2} Fitting}

In strained Mn$_3$NiN, eight distinct domain variants of the $\Gamma^{4g}$ phase exist [see Figs.~\ref{fig_4}(a)-(d) and ~\ref{fig_7}], each associated with a particular orientation of piezo-induced magnetic moment $\mathbf{M_P}$ [black arrows in Figs.~\ref{fig_4}(a)–\ref{fig_4}(d) and \ref{fig_7}]. For a given \textbf{H}, the relative domain populations are governed by the angle between $\mathbf{M_P}$ in a given domain and \textbf{H} (see Table~\ref{table1}). Domains for which $\mathbf{M_P}$ is more closely aligned with \textbf{H} are energetically favored and therefore more strongly populated, whereas domains with larger misalignment are comparatively suppressed. For simplicity, we are neglecting the effects of domain pinning and defects arising from slip planes \cite{Johnson2024TheFilms}. An analogous consideration also applies to Mn$_3$GaN, where $\mathbf{M_P}$ are rotated by 90° in the spin plane (see Fig.~\ref{fig_8}). Notably, domains within pairs (1,7), (2,8), (3,5) and (4,6) in Mn$_3$GaN share the same direction of $\mathbf{M_P}$ (see Table~\ref{table2} and Fig.~\ref{fig_8}).

In the zero-field case for Mn$_3$NiN, we assume an equal population of all eight domains, each occupying $100\,\% / 8 = 12.5\,\%$ of the sample volume. For an applied field +\textbf{H} (-\textbf{H}) at $\Theta=20$°, we parameterize the field-induced domain redistribution by increasing (decreasing) the volume fractions of domains~1 and~4 by $N^+$ and of domains~6 and~7 by $N^-$. Correspondingly, the volumes of domains~5 and~8 are decreased (increased) by $N^+$, and those of domains~2 and~3 are decreased (increased) by $N^-$. We treated $N^+$ and $N^-$ as fitting parameters in the weighted-domain average, subject to the constraint $0<N^-<N^+$. It reflects the assumption that domains with a larger projection of $\mathbf{M_P}$ onto \textbf{H} experience a larger increase in population. 

Pump-induced magnetic quenching was modeled by reducing the parameters $K$ and $Q$. For simplicity, we assumed that the pump modifies the real and imaginary parts of each parameter by the same relative amount. Since $K$ is linear in \textbf{M} whereas $Q$ is quadratic in \textbf{M}, we parameterized the quenching by a single factor $C$ with $0 < C < 1$, such that $K \rightarrow C K$ and $Q \rightarrow C^2 Q$, where the arrows represent the excitation. Non-magnetic pump-induced changes were modeled by modifying $\epsilon^{\left(0\right)}$, which captures the change of refractive index. Again for simplicity, we applied the same relative scaling to the real and imaginary parts of $\epsilon^{\left(0\right)}$, allowing the corresponding multiplicative factor to be either greater or smaller than unity \cite{DelFatti2000NonequilibriumMetals}.

In total, the model contains four fitting parameters used to reproduce the experimentally measured $\beta$-dependence: $N^+$, $N^-$, the change of $\epsilon^{\left(0\right)}$ (heating), and $C$, the change of $K$ and $Q$ (magnetic quenching). The best fit of the measured $\beta$-dependence [symbols in Fig.~\ref{fig_3}(c)] is obtained for $N^+ = 0.12\pm0.09\,\%$, $N^- = 0.11\pm0.9\,\%$, a $0.2\pm0.1\,\%$ increase of $\epsilon^{\left(0\right)}$, and $C=0.5\pm0.2$, which corresponds to a $50\pm20\,\%$ magnetic quenching.

To model the configuration with sample tilt $\Theta = 0$°, we set $N^+ = N^- = 0.11\,\%$ and used the same pump-induced changes as obtained from the fit at $\Theta = 20$°, namely 50\,\% magnetic quenching and a 0.2\,\% increase of $\epsilon^{\left(0\right)}$. The resulting $\beta$-dependencies are shown in Figs.~\ref{fig_4}(f)–\ref{fig_4}(h) as dashed blue and red curves.

\begin{table}
\caption{\label{table1}%
Directions of piezo-induced magnetic moment $\mathbf{M_P}$ in all 8 variants of $\Gamma^{4g}$, together with an angle between $\mathbf{M_P}$ and applied magnetic fields $\pm \mathbf{H}$ for $\Theta = 0$° and 20°. Cases with positive projection of $\mathbf{M_P}$ to \textbf{H} are marked in bold.}
\begin{ruledtabular}
\begin{tabular}{cccccc}
\multirow{3}{*}{Domain} & \multirow{3}{*}{\begin{tabular}[c]{@{}c@{}}$\mathbf{M_P}$\end{tabular}} & \multicolumn{4}{c}{Angle between $\mathbf{M_P}$ and} \\
 &  & \begin{tabular}[c]{@{}c@{}}+\textbf{H}\\ $\Theta = 0$°\end{tabular} & \begin{tabular}[c]{@{}c@{}}-\textbf{H}\\ $\Theta = 0$°\end{tabular} & \begin{tabular}[c]{@{}c@{}}+\textbf{H}\\ $\Theta = 20$°\end{tabular} & \begin{tabular}[c]{@{}c@{}}-\textbf{H}\\  $\Theta = 20$°\end{tabular} \\ \hline
1 & [-1,1,2] & \textbf{65.9°} & 114.1° & \textbf{48.5°} & 131.5° \\
2 & [-1,-1,2] & 114.1° & \textbf{65.9°} & 96° & \textbf{84°} \\
3 & [1,-1,2] & 114.1° & \textbf{65.9°} & 96° & \textbf{84°} \\
4 & [1,1,2] & \textbf{65.9°} & 114.1° & \textbf{48.5°} & 131.5° \\
5 & [1,-1,-2] & 114.1° & \textbf{65.9°} & 131.5° & \textbf{48.5°} \\
6 & [1,1,-2] & \textbf{65.9°} & 114.1° & \textbf{84°} & 96° \\
7 & [-1,1,-2] & \textbf{65.9°} & 114.1° & \textbf{84°} & 96° \\
8 & [-1,-1,-2] & 114.1° & \textbf{65.9°} & 131.5° & \textbf{48.5°}
\end{tabular}
\end{ruledtabular}
\end{table}

\begin{table}
\caption{\label{table2}%
Directions of piezo-induced magnetic moment $\mathbf{M_P}$ in all 8 variants of $\Gamma^{5g}$, together with an angle between $\mathbf{M_P}$ and applied magnetic fields $\pm \mathbf{H}$ for $\Theta = 0$° and 20°. Cases with positive projection of $\mathbf{M_P}$ to \textbf{H} are marked in bold.}
\begin{ruledtabular}
\begin{tabular}{cccccc}
\multirow{3}{*}{Domain} & \multirow{3}{*}{\begin{tabular}[c]{@{}c@{}}$\mathbf{M_P}$\end{tabular}} & \multicolumn{4}{c}{Angle between $\mathbf{M_P}$ and} \\
 &  & \begin{tabular}[c]{@{}c@{}}+\textbf{H}\\ $\Theta = 0$°\end{tabular} & \begin{tabular}[c]{@{}c@{}}-\textbf{H}\\ $\Theta = 0$°\end{tabular} & \begin{tabular}[c]{@{}c@{}}+\textbf{H}\\ $\Theta = 20$°\end{tabular} & \begin{tabular}[c]{@{}c@{}}-\textbf{H}\\  $\Theta = 20$°\end{tabular} \\ \hline
1, 7 & [1,1,0] & \textbf{45°} & 135° & \textbf{48.4°} & 131.6°  \\
2, 8 & [-1,1,0] & \textbf{45°} & 135° & \textbf{48.4°} & 131.6°  \\
3, 5 & [-1,-1,0] & 135° & \textbf{45°} & 131.6° &  \textbf{48.4°} \\
4, 6 & [1,-1,0] & 135° & \textbf{45°} & 131.6° &  \textbf{48.4°}
\end{tabular}
\end{ruledtabular}
\end{table}

To illustrate the model response for different quenching magnitudes, we fit the experimentally obtained $\beta$-dependence with various but fixed magnitudes of magnetic quenching. The resulting fits for both polarities of \textbf{H} are shown as curves in Figs.~\ref{fig_10}(a) and \ref{fig_10}(b) for fixed quenching magnitudes of 0.5\,\% and 100\,\%. Corresponding changes in domain populations were: $N^+ = 0.2\,\%$, $N^- = 0.05\,\%$  for 0.5\,\% magnetic quenching; $N^+ = 0.046\,\%$, $N^- = 0.044\,\%$ for 100\,\% magnetic quenching. In both cases, the change of $\epsilon^{\left(0\right)}$ was fixed to a 0.2\,\% increase.

In Mn$_3$GaN, domains~1, 2, 7, and~8 are equally populated and, moreover, are favored over the remaining domains for both $\Theta = 0$° and $\Theta = 20$° (see Table~\ref{table2}). As a consequence, at $\Theta = 0$° the net Voigt response almost completely vanishes (see red and blue curves in Fig.~\ref{fig_11}), because the contributions from individual domains are degenerate [see Fig.~\ref{fig_9}(b)]. At $\Theta = 20$°, the degeneracy is lifted at the level of the individual domain responses [see Fig.~\ref{fig_9}(d)], yielding a finite Voigt effect (see black and green curves in Fig.~\ref{fig_11}). Nevertheless, the combined signal generated by domains~1, 2, 7, and~8 equals the combined signal from the remaining domains. Therefore, even though the applied magnetic field probably redistributes the relative populations between these two domain groups, the total Voigt response remains unchanged. In the present geometry, the resulting Voigt contribution for Mn$_3$GaN is thus independent of the applied magnetic field.

\bibliography{references}

@article{Yan2019AFields,
    title = {{A piezoelectric, strain-controlled antiferromagnetic memory insensitive to magnetic fields}},
    year = {2019},
    journal = {Nature Nanotechnology},
    author = {Yan, Han and Feng, Zexin and Shang, Shunli and Wang, Xiaoning and Hu, Zexiang and Wang, Jinhua and Zhu, Zengwei and Wang, Hui and Chen, Zuhuang and Hua, Hui and Lu, Wenkuo and Wang, Jingmin and Qin, Peixin and Guo, Huixin and Zhou, Xiaorong and Leng, Zhaoguogang and Liu, Zikui and Jiang, Chengbao and Coey, Michael and Liu, Zhiqi},
    number = {2},
    month = {2},
    pages = {131--136},
    volume = {14},
    doi = {10.1038/s41565-018-0339-0},
    issn = {1748-3387}
}

@article{Kresse1993AbMetals,
    title = {{Ab initio molecular dynamics for liquid metals}},
    year = {1993},
    journal = {Physical Review B},
    author = {Kresse, G. and Hafner, J.},
    number = {1},
    month = {1},
    pages = {558--561},
    volume = {47},
    doi = {10.1103/PhysRevB.47.558},
    issn = {0163-1829}
}

@article{Gurung2019AnomalousAntiperovskites,
    title = {{Anomalous Hall conductivity of noncollinear magnetic antiperovskites}},
    year = {2019},
    journal = {Physical Review Materials},
    author = {Gurung, Gautam and Shao, Ding Fu and Paudel, Tula R. and Tsymbal, Evgeny Y.},
    number = {4},
    month = {4},
    volume = {3},
    publisher = {American Physical Society},
    doi = {10.1103/PhysRevMaterials.3.044409},
    issn = {24759953},
    arxivId = {1901.05040}
}

@article{Boldrin2019AnomalousFilms,
    title = {{Anomalous Hall effect in noncollinear antiferromagnetic Mn3NiN thin films}},
    year = {2019},
    journal = {Physical Review Materials},
    author = {Boldrin, David and Samathrakis, Ilias and Zemen, Jan and Mihai, Andrei and Zou, Bin and Johnson, Freya and Esser, Bryan D. and McComb, David W. and Petrov, Peter K. and Zhang, Hongbin and Cohen, Lesley F.},
    number = {9},
    month = {9},
    volume = {3},
    publisher = {American Physical Society},
    doi = {10.1103/PhysRevMaterials.3.094409},
    issn = {24759953},
    arxivId = {1902.04357}
}

@article{Beckert2023AnomalousFilms,
    title = {{Anomalous Nernst effect in Mn3NiN thin films}},
    year = {2023},
    journal = {Physical Review B},
    author = {Beckert, Sebastian and Godinho, João and Johnson, Freya and Kim{\'{a}}k, Jozef and Schmoranzerov{\'{a}}, Eva and Zemen, Jan and {\v{S}}ob{\'{a}}ň, Zbyněk and Olejn{\'{i}}k, Kamil and {\v{Z}}elezn{\'{y}}, Jakub and Wunderlich, Joerg and N{\v{e}}mec, Petr and Kriegner, Dominik and Thomas, Andy and Goennenwein, Sebastian T. B. and Cohen, Lesley F. and Reichlov{\'{a}}, Helena},
    number = {2},
    month = {7},
    pages = {024420},
    volume = {108},
    doi = {10.1103/PhysRevB.108.024420},
    issn = {2469-9950}
}

@article{Nemec2018AntiferromagneticOpto-spintronics,
    title = {{Antiferromagnetic opto-spintronics}},
    year = {2018},
    journal = {Nature Physics},
    author = {N{\v{e}}mec, P. and Fiebig, M. and Kampfrath, T. and Kimel, A. V.},
    number = {3},
    month = {3},
    pages = {229--241},
    volume = {14},
    doi = {10.1038/s41567-018-0051-x},
    issn = {1745-2473}
}

@article{Baltz2018AntiferromagneticSpintronics,
    title = {{Antiferromagnetic spintronics}},
    year = {2018},
    journal = {Reviews of Modern Physics},
    author = {Baltz, V. and Manchon, A. and Tsoi, M. and Moriyama, T. and Ono, T. and Tserkovnyak, Y.},
    number = {1},
    month = {2},
    pages = {015005},
    volume = {90},
    doi = {10.1103/RevModPhys.90.015005},
    issn = {0034-6861}
}

@article{Jungwirth2016AntiferromagneticSpintronics,
    title = {{Antiferromagnetic spintronics}},
    year = {2016},
    journal = {Nature Nanotechnology},
    author = {Jungwirth, T. and Marti, X. and Wadley, P. and Wunderlich, J.},
    number = {3},
    month = {3},
    pages = {231--241},
    volume = {11},
    doi = {10.1038/nnano.2016.18},
    issn = {1748-3387}
}

@misc{KimakArXiv:2601.07753,
    title = {{arXiv:2601.07753}},
    author = {Kim{\'{a}}k, J and Nerodilov{\'{a}}, M and Carva, K and Ghosh, S and {\v{Z}}elezn{\'{y}}, J and Ostatnick{\'{y}}, T and Zemen, J and Johnson, F and Boldrin, D and Rendell-Bhatti, F and Zou, B and Mihai, A P and Sun, X and Yu, F and Schmoranzerov{\'{a}}, E and N{\'{a}}dvorn{\'{i}}k, L and Cohen, L F and N{\v{e}}mec, P},
    url = {https://arxiv.org/abs/2601.07753},
    doi = {https://doi.org/10.48550/arXiv.2601.07753},
    arxivId = {2601.07753}
}

@article{Veis2018BandSpectroscopy,
    title = {{Band structure of CuMnAs probed by optical and photoemission spectroscopy}},
    year = {2018},
    journal = {Physical Review B},
    author = {Veis, M. and Min{\'{a}}r, J. and Steciuk, G. and Palatinus, L. and Rinaldi, C. and Cantoni, M. and Kriegner, D. and Tikui{\v{s}}is, K. K. and Hamrle, J. and Zahradn{\'{i}}k, M. and Anto{\v{s}}, R. and {\v{Z}}elezn{\'{y}}, J. and {\v{S}}mejkal, L. and Marti, X. and Wadley, P. and Campion, R. P. and Frontera, C. and Uhl{\'{i}}{\v{r}}ov{\'{a}}, K. and Duchoň, T. and Ku{\v{z}}el, P. and Nov{\'{a}}k, V. and Jungwirth, T. and V{\'{y}}born{\'{y}}, K.},
    number = {12},
    month = {3},
    pages = {125109},
    volume = {97},
    doi = {10.1103/PhysRevB.97.125109},
    issn = {2469-9950}
}

@article{Shi2016BaromagneticAnalysis,
    title = {{Baromagnetic Effect in Antiperovskite Mn3Ga0.95N0.94 by Neutron Powder Diffraction Analysis}},
    year = {2016},
    journal = {Advanced Materials},
    author = {Shi, Kewen and Sun, Ying and Yan, Jun and Deng, Sihao and Wang, Lei and Wu, Hui and Hu, Pengwei and Lu, Huiqing and Malik, Muhammad Imran and Huang, Qingzhen and Wang, Cong},
    number = {19},
    month = {5},
    pages = {3761--3767},
    volume = {28},
    doi = {10.1002/adma.201600310},
    issn = {0935-9648}
}

@article{Suzuki2017ClusterAntiferromagnets,
    title = {{Cluster multipole theory for anomalous Hall effect in antiferromagnets}},
    year = {2017},
    journal = {Physical Review B},
    author = {Suzuki, M.-T. and Koretsune, T. and Ochi, M. and Arita, R.},
    number = {9},
    month = {3},
    pages = {094406},
    volume = {95},
    doi = {10.1103/PhysRevB.95.094406},
    issn = {2469-9950}
}

@article{Zemen2023CollinearSpectroscopy,
    title = {{Collinear and noncollinear ferrimagnetic phases in Mn4N investigated by magneto-optical Kerr spectroscopy}},
    year = {2023},
    journal = {Journal of Applied Physics},
    author = {Zemen, Jan},
    number = {20},
    month = {11},
    volume = {134},
    publisher = {American Institute of Physics Inc.},
    doi = {10.1063/5.0170621},
    issn = {10897550}
}

@article{Liu2018ElectricalTemperature,
    title = {{Electrical switching of the topological anomalous Hall effect in a non-collinear antiferromagnet above room temperature}},
    year = {2018},
    journal = {Nature Electronics},
    author = {Liu, Z. Q. and Chen, H. and Wang, J. M. and Liu, J. H. and Wang, K. and Feng, Z. X. and Yan, H. and Wang, X. R. and Jiang, C. B. and Coey, J. M. D. and MacDonald, A. H.},
    number = {3},
    month = {3},
    pages = {172--177},
    volume = {1},
    doi = {10.1038/s41928-018-0040-1},
    issn = {2520-1131}
}

@article{Sultan2012Electron-Gd0001,
    title = {{Electron- and phonon-mediated ultrafast magnetization dynamics of Gd(0001)}},
    year = {2012},
    journal = {Physical Review B},
    author = {Sultan, Muhammad and Atxitia, Unai and Melnikov, Alexey and Chubykalo-Fesenko, Oksana and Bovensiepen, Uwe},
    number = {18},
    month = {5},
    pages = {184407},
    volume = {85},
    doi = {10.1103/PhysRevB.85.184407},
    issn = {1098-0121}
}

@article{Dudarev1998Electron-energy-lossStudy,
    title = {{Electron-energy-loss spectra and the structural stability of nickel oxide: An LSDA U study}},
    year = {1998},
    journal = {Physical Review B},
    author = {Dudarev, S. L. and Botton, G. A. and Savrasov, S. Y. and Humphreys, C. J. and Sutton, A. P.},
    number = {3},
    month = {1},
    pages = {1505--1509},
    volume = {57},
    doi = {10.1103/PhysRevB.57.1505},
    issn = {0163-1829}
}

@article{Abbas2025ExperimentalMgO001,
    title = {{Experimental and theoretical investigation of the crystalline surface, film, and interface properties of antiperovskite Mn3GaN grown by molecular beam epitaxy on MgO(001)}},
    year = {2025},
    journal = {Surfaces and Interfaces},
    author = {Abbas, Ali and Hernandez, Juan Carlos Moreno and Shrestha, Ashok and Russell, Daniel and Erickson, Tyler and Sun, Kai and Cocoletzi, Gregorio Hernandez and Yang, Fengyuan and Smith, Arthur R.},
    month = {5},
    pages = {106201},
    volume = {64},
    doi = {10.1016/j.surfin.2025.106201},
    issn = {24680230}
}

@article{Koopmans2010ExplainingDemagnetization,
    title = {{Explaining the paradoxical diversity of ultrafast laser-induced demagnetization}},
    year = {2010},
    journal = {Nature Materials 2010 9:3},
    author = {Koopmans, B. and Malinowski, G. and Dalla Longa, F. and Steiauf, D. and F{\"{a}}hnle, M. and Roth, T. and Cinchetti, M. and Aeschlimann, M.},
    number = {3},
    month = {12},
    pages = {259--265},
    volume = {9},
    publisher = {Nature Publishing Group},
    url = {https://www.nature.com/articles/nmat2593},
    doi = {10.1038/nmat2593},
    issn = {1476-4660}
}

@article{Sun1994Femtosecond-tunableGold,
    title = {{Femtosecond-tunable measurement of electron thermalization in gold}},
    year = {1994},
    journal = {Physical Review B},
    author = {Sun, C. K. and Vall{\'{e}}e, F. and Acioli, L. H. and Ippen, E. P. and Fujimoto, J. G.},
    number = {20},
    month = {11},
    pages = {15337},
    volume = {50},
    publisher = {American Physical Society},
    url = {https://journals.aps.org/prb/abstract/10.1103/PhysRevB.50.15337},
    doi = {10.1103/PhysRevB.50.15337},
    issn = {01631829},
    pmid = {9975886}
}

@article{Zemen2017FrustratedTheory,
    title = {{Frustrated magnetism and caloric effects in Mn-based antiperovskite nitrides: Ab initio theory}},
    year = {2017},
    journal = {Physical Review B},
    author = {Zemen, J. and Mendive-Tapia, E. and Gercsi, Z. and Banerjee, R. and Staunton, J. B. and Sandeman, K. G.},
    number = {18},
    month = {5},
    pages = {184438},
    volume = {95},
    doi = {10.1103/PhysRevB.95.184438},
    issn = {2469-9950}
}

@article{Matsunami2015GiantMn3GaN,
    title = {{Giant barocaloric effect enhanced by the frustration of the antiferromagnetic phase in Mn3GaN}},
    year = {2015},
    journal = {Nature Materials},
    author = {Matsunami, Daichi and Fujita, Asaya and Takenaka, Koshi and Kano, Mika},
    number = {1},
    month = {1},
    pages = {73--78},
    volume = {14},
    doi = {10.1038/nmat4117},
    issn = {1476-1122}
}

@article{Miwa2021GiantSemimetal,
    title = {{Giant Effective Damping of Octupole Oscillation in an Antiferromagnetic Weyl Semimetal}},
    year = {2021},
    journal = {Small Science},
    author = {Miwa, Shinji and Iihama, Satoshi and Nomoto, Takuya and Tomita, Takahiro and Higo, Tomoya and Ikhlas, Muhammad and Sakamoto, Shoya and Otani, YoshiChika and Mizukami, Shigemi and Arita, Ryotaro and Nakatsuji, Satoru},
    number = {5},
    month = {5},
    pages = {2000062},
    volume = {1},
    doi = {10.1002/smsc.202000062},
    issn = {2688-4046}
}

@article{Boldrin2018GiantMn3NiN,
    title = {{Giant Piezomagnetism in Mn3NiN}},
    year = {2018},
    journal = {ACS Applied Materials and Interfaces},
    author = {Boldrin, David and Mihai, Andrei P. and Zou, Bin and Zemen, Jan and Thompson, Ryan and Ware, Ecaterina and Neamtu, Bogdan V. and Ghivelder, Luis and Esser, Bryan and McComb, David W. and Petrov, Peter and Cohen, Lesley F.},
    number = {22},
    month = {6},
    pages = {18863--18868},
    volume = {10},
    publisher = {American Chemical Society},
    doi = {10.1021/acsami.8b03112},
    issn = {19448252},
    pmid = {29726252}
}

@article{Guo2020GiantMetal,
    title = {{Giant Piezospintronic Effect in a Noncollinear Antiferromagnetic Metal}},
    year = {2020},
    journal = {Advanced Materials},
    author = {Guo, Huixin and Feng, Zexin and Yan, Han and Liu, Jiuzhao and Zhang, Jia and Zhou, Xiaorong and Qin, Peixin and Cai, Jialin and Zeng, Zhongming and Zhang, Xin and Wang, Xiaoning and Chen, Hongyu and Wu, Haojiang and Jiang, Chengbao and Liu, Zhiqi},
    number = {26},
    month = {7},
    pages = {2002300},
    volume = {32},
    doi = {10.1002/adma.202002300},
    issn = {0935-9648}
}

@article{Schmoranzerova2022GiantExperiments,
    title = {{Giant quadratic magneto-optical response of thin thin Y3Fe5O12 films for sensitive magnetometry experiments}},
    year = {2022},
    journal = {Physical Review B},
    author = {Schmoranzerov{\'{a}}, E. and Ostatnick{\'{y}}, T. and Kim{\'{a}}k, J. and Kriegner, D. and Reichlov{\'{a}}, H. and Schlitz, R. and Baďura, A. and {\v{S}}ob{\'{a}}ň, Z. and M{\"{u}}nzenberg, M. and Jakob, G. and Guo, E.-J. and Kl{\"{a}}ui, M. and N{\v{e}}mec, P.},
    number = {10},
    month = {9},
    pages = {104434},
    volume = {106},
    doi = {10.1103/PhysRevB.106.104434},
    issn = {2469-9950}
}

@article{Pankratova2022Heat-conservingNickel,
    title = {{Heat-conserving three-temperature model for ultrafast demagnetization in nickel}},
    year = {2022},
    journal = {Physical Review B},
    author = {Pankratova, M. and Miranda, I. P. and Thonig, D. and Pereiro, M. and Sj{\"{o}}qvist, E. and Delin, A. and Eriksson, O. and Bergman, A.},
    number = {17},
    month = {11},
    pages = {174407},
    volume = {106},
    doi = {10.1103/PhysRevB.106.174407},
    issn = {2469-9950}
}

@article{Johnson2022IdentifyingMicroscopy,
    title = {{Identifying the octupole antiferromagnetic domain orientation in Mn3NiN by scanning anomalous Nernst effect microscopy}},
    year = {2022},
    journal = {Applied Physics Letters},
    author = {Johnson, F. and Kim{\'{a}}k, J. and Zemen, J. and {\v{S}}ob{\'{a}}ň, Z. and Schmoranzerov{\'{a}}, E. and Godinho, J. and N{\v{e}}mec, P. and Beckert, S. and Reichlov{\'{a}}, H. and Boldrin, D. and Wunderlich, J. and Cohen, L. F.},
    number = {23},
    month = {6},
    pages = {232402},
    volume = {120},
    doi = {10.1063/5.0091257},
    issn = {0003-6951}
}

@article{Reichlova2019ImagingMn3Sn,
    title = {{Imaging and writing magnetic domains in the non-collinear antiferromagnet Mn3Sn}},
    year = {2019},
    journal = {Nature Communications},
    author = {Reichlova, Helena and Janda, Tomas and Godinho, Joao and Markou, Anastasios and Kriegner, Dominik and Schlitz, Richard and Zelezny, Jakub and Soban, Zbynek and Bejarano, Mauricio and Schultheiss, Helmut and Nemec, Petr and Jungwirth, Tomas and Felser, Claudia and Wunderlich, Joerg and Goennenwein, Sebastian T. B.},
    number = {1},
    month = {11},
    pages = {5459},
    volume = {10},
    doi = {10.1038/s41467-019-13391-z},
    issn = {2041-1723}
}

@article{Surynek2020InvestigationExperiment,
    title = {{Investigation of magnetic anisotropy and heat dissipation in thin films of compensated antiferromagnet CuMnAs by pump–probe experiment}},
    year = {2020},
    journal = {Journal of Applied Physics},
    author = {Sur{\'{y}}nek, M. and Saidl, V. and Ka{\v{s}}par, Z. and Nov{\'{a}}k, V. and Campion, R. P. and Wadley, P. and N{\v{e}}mec, P.},
    number = {23},
    month = {6},
    pages = {233904},
    volume = {127},
    doi = {10.1063/5.0006185},
    issn = {0021-8979}
}

@incollection{Chubykalo-Fesenko2020Landau-Lifshitz-BlochTransition,
    title = {{Landau-Lifshitz-Bloch Approach for Magnetization Dynamics Close to Phase Transition}},
    year = {2020},
    booktitle = {Handbook of Materials Modeling},
    author = {Chubykalo-Fesenko, Oksana and Nieves, Pablo},
    pages = {867--893},
    publisher = {Springer International Publishing},
    address = {Cham},
    doi = {10.1007/978-3-319-44677-6{\_}72}
}

@article{Nayak2016LargeMn3Ge,
    title = {{Large anomalous Hall effect driven by a nonvanishing Berry curvature in the noncolinear antiferromagnet Mn3Ge}},
    year = {2016},
    journal = {Science Advances},
    author = {Nayak, Ajaya K. and Fischer, Julia Erika and Sun, Yan and Yan, Binghai and Karel, Julie and Komarek, Alexander C. and Shekhar, Chandra and Kumar, Nitesh and Schnelle, Walter and K{\"{u}}bler, Jürgen and Felser, Claudia and Parkin, Stuart S. P.},
    number = {4},
    month = {4},
    volume = {2},
    doi = {10.1126/sciadv.1501870},
    issn = {2375-2548}
}

@article{Nakatsuji2015LargeTemperature,
    title = {{Large anomalous Hall effect in a non-collinear antiferromagnet at room temperature}},
    year = {2015},
    journal = {Nature},
    author = {Nakatsuji, Satoru and Kiyohara, Naoki and Higo, Tomoya},
    number = {7577},
    month = {11},
    pages = {212--215},
    volume = {527},
    doi = {10.1038/nature15723},
    issn = {0028-0836}
}

@article{Higo2018LargeMetal,
    title = {{Large magneto-optical Kerr effect and imaging of magnetic octupole domains in an antiferromagnetic metal}},
    year = {2018},
    journal = {Nature Photonics},
    author = {Higo, Tomoya and Man, Huiyuan and Gopman, Daniel B. and Wu, Liang and Koretsune, Takashi and van ’t Erve, Olaf M. J. and Kabanov, Yury P. and Rees, Dylan and Li, Yufan and Suzuki, Michi-To and Patankar, Shreyas and Ikhlas, Muhammad and Chien, C. L. and Arita, Ryotaro and Shull, Robert D. and Orenstein, Joseph and Nakatsuji, Satoru},
    number = {2},
    month = {2},
    pages = {73--78},
    volume = {12},
    doi = {10.1038/s41566-017-0086-z},
    issn = {1749-4885}
}

@article{Feng2015LargePt,
    title = {{Large magneto-optical Kerr effect in noncollinear antiferromagnets Mn3X (X = Rh, Ir, or Pt)}},
    year = {2015},
    journal = {Physical Review B},
    author = {Feng, Wanxiang and Guo, Guang-Yu and Zhou, Jian and Yao, Yugui and Niu, Qian},
    number = {14},
    month = {10},
    pages = {144426},
    volume = {92},
    doi = {10.1103/PhysRevB.92.144426},
    issn = {1098-0121}
}

@article{Zhao2021LargeMn3Sn,
    title = {{Large ultrafast-modulated Voigt effect in noncollinear antiferromagnet Mn3Sn}},
    year = {2021},
    journal = {Nature Communications},
    author = {Zhao, H. C. and Xia, H. and Hu, S. and Lv, Y. Y. and Zhao, Z. R. and He, J. and Liang, E. and Ni, G. and Chen, L. Y. and Qiu, X. P. and Zhou, S. M. and Zhao, H. B.},
    number = {1},
    month = {9},
    pages = {5266},
    volume = {12},
    doi = {10.1038/s41467-021-25654-9},
    issn = {2041-1723}
}

@article{Zheng2018Magneto-opticalFilms,
    title = {{Magneto-optical probe of ultrafast spin dynamics in antiferromagnetic CoO thin films}},
    year = {2018},
    journal = {Physical Review B},
    author = {Zheng, Z. and Shi, J. Y. and Li, Q. and Gu, T. and Xia, H. and Shen, L. Q. and Jin, F. and Yuan, H. C. and Wu, Y. Z. and Chen, L. Y. and Zhao, H. B.},
    number = {13},
    month = {10},
    pages = {134409},
    volume = {98},
    doi = {10.1103/PhysRevB.98.134409},
    issn = {2469-9950}
}

@article{Rimmler2024Non-collinearSpintronics,
    title = {{Non-collinear antiferromagnetic spintronics}},
    year = {2024},
    journal = {Nature Reviews Materials},
    author = {Rimmler, Berthold H. and Pal, Banabir and Parkin, Stuart S. P.},
    number = {2},
    month = {8},
    pages = {109--127},
    volume = {10},
    doi = {10.1038/s41578-024-00706-w},
    issn = {2058-8437}
}

@article{DelFatti2000NonequilibriumMetals,
    title = {{Nonequilibrium electron dynamics in noble metals}},
    year = {2000},
    journal = {Physical Review B},
    author = {Del Fatti, N. and Voisin, C. and Achermann, M. and Tzortzakis, S. and Christofilos, D. and Vall{\'{e}}e, F.},
    number = {24},
    month = {6},
    pages = {16956--16966},
    volume = {61},
    doi = {10.1103/PhysRevB.61.16956},
    issn = {0163-1829}
}

@article{DelFatti1998NonequilibriumFilms,
    title = {{Nonequilibrium Electron Interactions in Metal Films}},
    year = {1998},
    journal = {Physical Review Letters},
    author = {Del Fatti, N. and Bouffanais, R. and Vall{\'{e}}e, F. and Flytzanis, C.},
    number = {4},
    month = {7},
    pages = {922--925},
    volume = {81},
    doi = {10.1103/PhysRevLett.81.922},
    issn = {0031-9007}
}

@article{Chen2023Octupole-drivenJunction,
    title = {{Octupole-driven magnetoresistance in an antiferromagnetic tunnel junction}},
    year = {2023},
    journal = {Nature},
    author = {Chen, Xianzhe and Higo, Tomoya and Tanaka, Katsuhiro and Nomoto, Takuya and Tsai, Hanshen and Idzuchi, Hiroshi and Shiga, Masanobu and Sakamoto, Shoya and Ando, Ryoya and Kosaki, Hidetoshi and Matsuo, Takumi and Nishio-Hamane, Daisuke and Arita, Ryotaro and Miwa, Shinji and Nakatsuji, Satoru},
    number = {7944},
    month = {1},
    pages = {490--495},
    volume = {613},
    doi = {10.1038/s41586-022-05463-w},
    issn = {0028-0836}
}

@article{Saidl2017OpticalAntiferromagnet,
    title = {{Optical determination of the N{\'{e}}el vector in a CuMnAs thin-film antiferromagnet}},
    year = {2017},
    journal = {Nature Photonics},
    author = {Saidl, V. and N{\v{e}}mec, P. and Wadley, P. and Hills, V. and Campion, R. P. and Nov{\'{a}}k, V. and Edmonds, K. W. and Maccherozzi, F. and Dhesi, S. S. and Gallagher, B. L. and Troj{\'{a}}nek, F. and Kune{\v{s}}, J. and {\v{Z}}elezn{\'{y}}, J. and Mal{\'{y}}, P. and Jungwirth, T.},
    number = {2},
    month = {2},
    pages = {91--96},
    volume = {11},
    doi = {10.1038/nphoton.2016.255},
    issn = {1749-4885}
}

@incollection{Visnovsky2006OpticsNanostructures,
    title = {{Optics in Magnetic Multilayers and Nanostructures}},
    year = {2006},
    author = {Vi{\v{s}}ňovsk{\'{y}}, {\v{S}}},
    chapter = {3},
    publisher = {CRC Press},
    address = {Boca Raton, FL}
}

@article{Zemen2017PiezomagnetismNitrides,
    title = {{Piezomagnetism as a counterpart of the magnetovolume effect in magnetically frustrated Mn-based antiperovskite nitrides}},
    year = {2017},
    journal = {Physical Review B},
    author = {Zemen, J. and Gercsi, Z. and Sandeman, K. G.},
    number = {2},
    month = {7},
    volume = {96},
    publisher = {American Physical Society},
    doi = {10.1103/PhysRevB.96.024451},
    issn = {24699969}
}

@article{Ishino2018PreparationCompositions,
    title = {{Preparation and evaluation of Mn3GaN1-x thin films with controlled N compositions}},
    year = {2018},
    journal = {AIP Advances},
    author = {Ishino, Sunao and So, Jongmin and Goto, Hirotaka and Hajiri, Tetsuya and Asano, Hidefumi},
    number = {5},
    month = {5},
    pages = {056312},
    volume = {8},
    doi = {10.1063/1.5007333},
    issn = {2158-3226}
}

@article{Wohlrath2025QuadraticAnisotropies,
    title = {{Quadratic magneto-optical Kerr effect spectroscopy: polarization variation method for investigation of magnetic and magneto-optical anisotropies}},
    year = {2025},
    journal = {Journal of Physics D: Applied Physics},
    author = {Wohlrath, V and Sadeghi, Z and Kim{\'{a}}k, J and Hovo{\v{r}}{\'{a}}kov{\'{a}}, K and Kuba{\v{s}}{\v{c}}{\'{i}}k, P and Schmoranzerov{\'{a}}, E and N{\'{a}}dvorn{\'{i}}k, L and Troj{\'{a}}nek, F and N{\v{e}}mec, P and Ostatnick{\'{y}}, T},
    number = {15},
    month = {4},
    pages = {155001},
    volume = {58},
    doi = {10.1088/1361-6463/adb760},
    issn = {0022-3727}
}

@article{You2018RevealingSpectroscopies,
    title = {{Revealing the Nature of the Ultrafast Magnetic Phase Transition in Ni by Correlating Extreme Ultraviolet Magneto-Optic and Photoemission Spectroscopies}},
    year = {2018},
    journal = {Physical Review Letters},
    author = {You, Wenjing and Tengdin, Phoebe and Chen, Cong and Shi, Xun and Zusin, Dmitriy and Zhang, Yingchao and Gentry, Christian and Blonsky, Adam and Keller, Mark and Oppeneer, Peter M. and Kapteyn, Henry and Tao, Zhensheng and Murnane, Margaret},
    number = {7},
    month = {8},
    pages = {077204},
    volume = {121},
    publisher = {American Physical Society},
    url = {https://journals.aps.org/prl/abstract/10.1103/PhysRevLett.121.077204},
    doi = {10.1103/PhysRevLett.121.077204},
    issn = {10797114},
    pmid = {30169091}
}

@article{Shim2020RoleMultilayer,
    title = {{Role of non-thermal electrons in ultrafast spin dynamics of ferromagnetic multilayer}},
    year = {2020},
    journal = {Scientific Reports},
    author = {Shim, Je-Ho and Syed, Akbar Ali and Kim, Jea-Il and Piao, Hong-Guang and Lee, Sang-Hyuk and Park, Seung-Young and Choi, Yeon Suk and Lee, Kyung Min and Kim, Hyun-Joong and Jeong, Jong-Ryul and Hong, Jung-Il and Kim, Dong Eon and Kim, Dong-Hyun},
    number = {1},
    month = {4},
    pages = {6355},
    volume = {10},
    doi = {10.1038/s41598-020-63452-3},
    issn = {2045-2322}
}

@article{Qin2023Room-temperatureJunction,
    title = {{Room-temperature magnetoresistance in an all-antiferromagnetic tunnel junction}},
    year = {2023},
    journal = {Nature},
    author = {Qin, Peixin and Yan, Han and Wang, Xiaoning and Chen, Hongyu and Meng, Ziang and Dong, Jianting and Zhu, Meng and Cai, Jialin and Feng, Zexin and Zhou, Xiaorong and Liu, Li and Zhang, Tianli and Zeng, Zhongming and Zhang, Jia and Jiang, Chengbao and Liu, Zhiqi},
    number = {7944},
    month = {1},
    pages = {485--489},
    volume = {613},
    doi = {10.1038/s41586-022-05461-y},
    issn = {0028-0836}
}

@article{Johnson2023Room-temperatureSignatures,
    title = {{Room-temperature weak collinear ferrimagnet with symmetry-driven large intrinsic magneto-optic signatures}},
    year = {2023},
    journal = {Physical Review B},
    author = {Johnson, F. and Z{\'{a}}zvorka, J. and Beran, L. and Boldrin, D. and Cohen, L. F. and Zemen, J. and Veis, M.},
    number = {1},
    month = {1},
    volume = {107},
    publisher = {American Physical Society},
    doi = {10.1103/PhysRevB.107.014404},
    issn = {24699969},
    arxivId = {2111.13498}
}

@article{Lee2025Spin-torque-drivenMn3Sn,
    title = {{Spin-torque-driven gigahertz magnetization dynamics in the non-collinear antiferromagnet Mn3Sn}},
    year = {2025},
    journal = {Nature Nanotechnology},
    author = {Lee, Won-Bin and Hwang, Seongmun and Ko, Hye-Won and Park, Byong-Guk and Lee, Kyung-Jin and Choi, Gyung-Min},
    number = {4},
    month = {4},
    pages = {487--493},
    volume = {20},
    doi = {10.1038/s41565-025-01859-7},
    issn = {1748-3387}
}

@article{Guo2025SpinPolarizedSpintronics,
    title = {{Spin‐Polarized Antiferromagnets for Spintronics}},
    year = {2025},
    journal = {Advanced Materials},
    author = {Guo, Zhenzhou and Wang, Xiaotian and Wang, Wenhong and Zhang, Gang and Zhou, Xiaodong and Cheng, Zhenxiang},
    number = {36},
    month = {9},
    pages = {2505779},
    volume = {37},
    doi = {10.1002/adma.202505779},
    issn = {0935-9648}
}

@article{Tesarova2014SystematicGaMnAs,
    title = {{Systematic study of magnetic linear dichroism and birefringence in (Ga,Mn)As}},
    year = {2014},
    journal = {Physical Review B},
    author = {Tesa{\v{r}}ov{\'{a}}, N. and Ostatnick{\'{y}}, T. and Nov{\'{a}}k, V. and Olejn{\'{i}}k, K. and {\v{S}}ubrt, J. and Reichlov{\'{a}}, H. and Ellis, C. T. and Mukherjee, A. and Lee, J. and Sipahi, G. M. and Sinova, J. and Hamrle, J. and Jungwirth, T. and N{\v{e}}mec, P. and {\v{C}}erne, J. and V{\'{y}}born{\'{y}}, K.},
    number = {8},
    month = {2},
    pages = {085203},
    volume = {89},
    doi = {10.1103/PhysRevB.89.085203},
    issn = {1098-0121}
}

@article{Roth2012TemperatureMechanism,
    title = {{Temperature Dependence of Laser-Induced Demagnetization in Ni: A Key for Identifying the Underlying Mechanism}},
    year = {2012},
    journal = {Physical Review X},
    author = {Roth, T. and Schellekens, A. J. and Alebrand, S. and Schmitt, O. and Steil, D. and Koopmans, B. and Cinchetti, M. and Aeschlimann, M.},
    number = {2},
    month = {5},
    pages = {021006},
    volume = {2},
    publisher = {American Physical Society},
    url = {https://journals.aps.org/prx/abstract/10.1103/PhysRevX.2.021006},
    doi = {10.1103/PhysRevX.2.021006},
    issn = {21603308}
}

@article{Boldrin2019TheFilms,
    title = {{The Biaxial Strain Dependence of Magnetic Order in Spin Frustrated Mn3NiN Thin Films}},
    year = {2019},
    journal = {Advanced Functional Materials},
    author = {Boldrin, David and Johnson, Freya and Thompson, Ryan and Mihai, Andrei P. and Zou, Bin and Zemen, Jan and Griffiths, Jack and Gubeljak, Patrik and Ormandy, Kristian L. and Manuel, Pascal and Khalyavin, Dmitry D. and Ouladdiaf, Bachir and Qureshi, Navid and Petrov, Peter and Branford, Will and Cohen, Lesley F.},
    number = {40},
    month = {10},
    volume = {29},
    publisher = {Wiley-VCH Verlag},
    doi = {10.1002/adfm.201902502},
    issn = {16163028}
}

@article{Johnson2024TheFilms,
    title = {{The Impact of Local Strain Fields in Noncollinear Antiferromagnetic Films}},
    year = {2024},
    journal = {Advanced Materials},
    author = {Johnson, Freya and Rendell-Bhatti, Frederic and Esser, Bryan D. and Hussey, Aisling and McComb, David W. and Zemen, Jan and Boldrin, David and Cohen, Lesley F.},
    number = {27},
    month = {7},
    volume = {36},
    publisher = {John Wiley and Sons Inc},
    doi = {10.1002/adma.202401180},
    issn = {15214095}
}

@article{Lukashev2008TheoryAntiperovskites,
    title = {{Theory of the piezomagnetic effect in Mn-based antiperovskites}},
    year = {2008},
    journal = {Physical Review B},
    author = {Lukashev, Pavel and Sabirianov, Renat F. and Belashchenko, Kirill},
    number = {18},
    month = {11},
    pages = {184414},
    volume = {78},
    doi = {10.1103/PhysRevB.78.184414},
    issn = {1098-0121}
}

@article{Gong2023UltrafastMagnet,
    title = {{Ultrafast demagnetization dynamics in the epitaxial FeGe(111) film chiral magnet}},
    year = {2023},
    journal = {Physical Review B},
    author = {Gong, Zizhao and Zhang, Wei and Liu, Jianan and Xie, Zongkai and Yang, Xu and Tang, Jin and Du, Haifeng and Li, Na and Zhang, Xiangqun and He, Wei and Cheng, Zhao-hua},
    number = {14},
    month = {4},
    pages = {144429},
    volume = {107},
    doi = {10.1103/PhysRevB.107.144429},
    issn = {2469-9950}
}

@article{Chen2025UltrafastProgress,
    title = {{Ultrafast demagnetization in ferromagnetic materials: Origins and progress}},
    year = {2025},
    journal = {Physics Reports},
    author = {Chen, Xiaowen and Adam, Roman and B{\"{u}}rgler, Daniel E. and Wang, Fangzhou and Lu, Zhenyan and Pan, Lining and Heidtfeld, Sarah and Greb, Christian and Liu, Meihong and Liu, Qingfang and Wang, Jianbo and Schneider, Claus M. and Cao, Derang},
    month = {1},
    pages = {1--63},
    volume = {1102},
    doi = {10.1016/j.physrep.2024.10.008},
    issn = {03701573}
}

@article{Atxitia2011UltrafastModel,
    title = {{Ultrafast magnetization dynamics rates within the Landau-Lifshitz-Bloch model}},
    year = {2011},
    journal = {Physical Review B},
    author = {Atxitia, U. and Chubykalo-Fesenko, O.},
    number = {14},
    month = {10},
    pages = {144414},
    volume = {84},
    doi = {10.1103/PhysRevB.84.144414},
    issn = {1098-0121}
}

@article{Kirilyuk2010UltrafastOrder,
    title = {{Ultrafast optical manipulation of magnetic order}},
    year = {2010},
    journal = {Reviews of Modern Physics},
    author = {Kirilyuk, Andrei and Kimel, Alexey V. and Rasing, Theo},
    number = {3},
    month = {9},
    pages = {2731--2784},
    volume = {82},
    publisher = {American Physical Society},
    issn = {15390756}
}

@article{Song2025UltrafastAntiferromagnets,
    title = {{Ultrafast Spin Current Excitation and Controlled Terahertz Radiation from Noncollinear Antiferromagnets}},
    year = {2025},
    journal = {Advanced Optical Materials},
    author = {Song, Yiwen and Lin, Dennis J. X. and Hu, Shanshan and Li, Ziyang and Zhang, Jiali and Lim, Bee Chun and Ko, Hnin Yu Yu and Chen, Shaohai and Ho, Pin and Jin, Qingyuan and Zhang, Zongzhi},
    number = {17},
    month = {6},
    pages = {2500210},
    volume = {13},
    doi = {10.1002/adom.202500210},
    issn = {2195-1071}
}

@article{Kim2009UltrafastFilm,
    title = {{Ultrafast spin demagnetization by nonthermal electrons of TbFe alloy film}},
    year = {2009},
    journal = {Applied Physics Letters},
    author = {Kim, Ji-Wan and Lee, Kyeong-Dong and Jeong, Jae-Woo and Shin, Sung-Chul},
    number = {19},
    month = {5},
    pages = {192506},
    volume = {94},
    doi = {10.1063/1.3130743},
    issn = {0003-6951}
}

@article{Beaurepaire1996UltrafastNickel,
    title = {{Ultrafast Spin Dynamics in Ferromagnetic Nickel}},
    year = {1996},
    journal = {Physical Review Letters},
    author = {Beaurepaire, E. and Merle, J.-C. and Daunois, A. and Bigot, J.-Y.},
    number = {22},
    month = {5},
    pages = {4250--4253},
    volume = {76},
    doi = {10.1103/PhysRevLett.76.4250},
    issn = {0031-9007}
}

\end{document}